# Two New Theories for the Current Charge Relativity and the Electric Origin of the Magnetic Force Between Two Filamentary Current Elements

*Waseem Ghassan Tahseen Shadid*

***Abstract*—This paper presents two new theories and a new current representation to explain the magnetic force between two filamentary current elements as a result of electric force interactions between current charges. The first theory states that a current has an electric charge relative to its moving observer. The second theory states that the magnetic force is an electric force in origin. The new current representation characterizes a current as equal amounts of positive and negative point charges moving in opposite directions at the speed of light. Previous work regarded electricity and magnetism as different aspects of the same subject. One effort was made by J.O. Johnson to unify the origin of electricity and magnetism, but this effort yielded a formula that is unequal to the well-known magnetic force law. The explanation provided for the magnetic force depends on three factors: (1) representing the electric current as charges moving at the speed of light, (2) considering the relative velocity between moving charges, and (3) analyzing the electric field spreading in the space due to the movement of charges inside current elements. The electric origin of the magnetic force is proved by deriving the magnetic force law and Biot-Savart law using the electric force law. This work is helpful for unifying the concepts of magnetism and electricity.**

***Index Terms*—Magnetic force, Electric force, Charge, Current, Relativity, Biot-Savart law, Electric field, Speed of light, Current filament.**

## I. Introduction

SINCE the invention of the voltaic cell in the early 19th century, many experiments have been conducted to study the force produced by two constant currents in loops; this force has been considered a new force and is different from the force produced by electrostatic charges [1]. In [2], Shadowitz provided a brief historical review of magnetism. The origin of the word 'magnetism' is a region in Asia Minor called Magnesia that has stones with the property of attracting similar small stones. The properties of magnetism were studied by William Gilbert in the 16th century. In the 19th century, Ft. C. Oersted discovered that a wire carrying electric current produces a force that affects a magnetized compass needle. This force was considered different in origin from the electrostatic force between charges because the current-carrying wires are intrinsically charge-neutral [1]. Years after Oersted's work, M. Faraday found a connection between electricity and magnetism through his discovery of electromagnetic induction. Then, Maxwell published his studies on the electromagnetic nature of light, followed by Hertz's work on the transmission and detection of electromagnetic waves. Since then, electricity and magnetism have been treated more and more as different aspects of the same subject [3], [4], [5]. In 1997, an effort was made by J.O. Johnson [6] to unify the origin of electricity and magnetism. However, the work ended up yielding a formula that is unequal to the well-known magnetic force law.

This work provides an explanation of the magnetic force between two filamentary current elements as a result of electric force interactions between current charges. This work introduces a new current representation and two new theories to explain the electric origin of magnetic force. The new current representation characterizes current as equal amounts of positive and negative point charges moving in opposite directions at the speed of light. This representation is referred to as the light-speed current representation. One new theory is introduced to show that a current has a relative charge with respect to its observer. A current's relative charge is zero, negative or positive depending on the motion of its observer. The second theory is introduced to show that the magnetic force between two filamentary current elements is a result of electric force interaction between current charges. As a proof, the exact formulas for the magnetic force law between two current elements and the Biot–Savart law are derived using the electric force law between electric charges. This work is developed to assist in unifying the concepts of electricity and magnetism. This unification may have important engineering benefits that may allow engineers to enhance the electrical properties for materials and to design new algorithms for computational electromagnetism.

There are twelve postulates and assumptions provided by [7], [8], [1] that are adopted by this work:

1) Relativity Principle (RP): All inertial frames are completely equivalent for the laws of physics. (gravity is not accounted for here and is negligible),
2) The speed of light in free space is constant and independent of its source and its receiver,
3) The charge of an electron or proton is mathematically considered herein to be a point charge, does not have a finite size, is massless, and is able to move at the speed of light,
4) An electric point charge emanates an electric field around it. This electric field spreads through the space equally in all directions, falling off in intensity at $1/r^2$,
5) Velocities between interacting objects are relative,
6) The electric field of a moving charge spreads through all of the space of its inertial frame at the speed of

Waseem G. T. Shadid address: P.O. Box 621564 Charlotte, North Carolina, USA, 28262 e-mail: wshadid78@gmail.com.

light, thus having an instant reaction with a charge that the field is in contact with. The acceleration of a charge creates a new velocity that changes the electric field that spreads out over the new inertial frame at the speed of light,

7) The charge value, $q$, is invariant from one inertial frame to another,
8) A positive sign on the overall electric/magnetic force represents repulsion, while a negative sign represents attraction,
9) A negative sign must be entered into the equations for negative charges, such as electrons. A positive sign may be entered into the equations for positive charges, such as protons. This makes the direction of the overall force appear correctly, as in 8 above,
10) Two objects cannot occupy the same position in the same space at the same time,
11) The amount and direction of a constant current flowing in a current element are constants and independent of its receiver,
12) Charges of a filamentary current are enforced to move freely along the filamentary curve only, without affecting it by any force and not permitted to leave it.

These postulates and assumptions are used to construct the concepts and theories developed in this work.

## II. Background

This section provides background information that is needed to understand the terminology of this work and to interpret the results. The section consists of an overview of four topics: (1) filamentary current, (2) infinitesimal current charge distance, (3) the Biot-Savart law, and (4) the magnetic force law between two current elements. This paper proposes new theories that integrate work from these topics to provide an explanation for the magnetic force using the concept of electric force.

### A. Filamentary Current

The current is filamentary when it flows through a long, very thin, conducting wire, where the charges are enforced to move freely along the filamentary curve only and are not permitted to leave it [2]. Thus, if a force is applied on a charge at point $\vec{p} = (x, y, z)$ along the direction of the filamentary current element, i.e., the differential element, the charge at that point moves along the filamentary curve direction without affecting the current element. If the applied force on the charge at point $\vec{p}$ is perpendicular to the filamentary current element at that point, the charge pushes the current element by that force because it cannot leave it.

### B. Infinitesimal Current Charge Distance

An infinitesimal current charge distance indicates that there is an infinitesimal distance between the centers of the moving positive and the negative charges that form a current in a filamentary current element. It is impossible for positive and negative charges to occupy the same position in the same space at the same time [8], as shown in figure (1) . The average net charge density in any region of a current at any time is zero [1]. The zero net charge of a current region indicates that the same amounts of positive and negative charges are present in that region, but they are either all moving or only some of them are moving within that region to generate the current. Figure (2) shows a model of how the positive and negative charges are estimated to be located in an infinitesimal current region $dV$ at moment $t$. The current in this model is assumed to be generated by the same amounts of positive and negative charges that are moving at the same speed but in opposite directions. The current propagates from right to left. The negative charge is on the left side of the region and is moving to the right side with speed $|\vec{v}|$, while the positive charge is on the right side and moving to the left side with speed $|\vec{v}|$. Thus, there is an infinitesimal distance between the centers of the positive and negative charges.

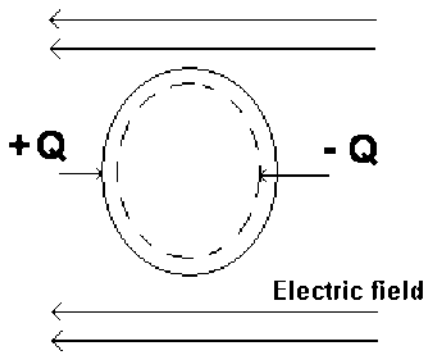

Figure 1: Illustrates how it is impossible for a positive charge and a negative charge to occupy the same location at the same time.

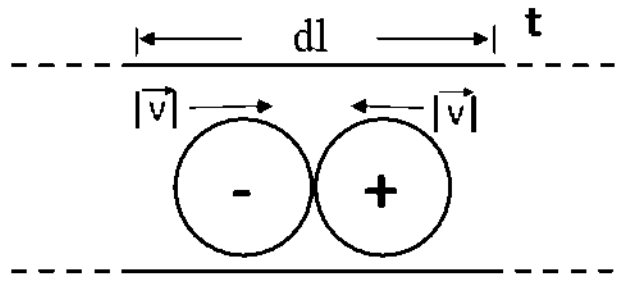

Figure 2: Shows the locations of charges and their directions of motion in an infinitesimal current region $dV$ at moment $t$. The current is assumed to be generated by the same amounts of positive and negative charges that are moving at the same speed but in opposite directions.

### C. Biot-Savart Law

The Biot-Savart law is an equation that quantifies the relationship between the electric current and the magnetic field it produces [9]. A magnetic field is a region in which the magnetic force is observed for a magnet source, e.g., a wire carrying current. The Biot-Savart law is defined as in equation (1) [10], [11].

$$\overrightarrow{dB(\vec{r})} = \frac{\mu}{4\pi} \frac{I}{|\vec{r}|^2} \overrightarrow{dl} \times \vec{a_r} \tag{1}$$

where $\overrightarrow{dB(\vec{r})}$ is the magnetic field resulting from an infinitesimal current element, i.e., a small segment of a current-carrying filament, at vector distance $\vec{r}$ from the current element to the field point. $\mu$ is the magnetic permeability. $\overrightarrow{dl}$ is the infinitesimal length of the element carrying current $I$. $\vec{a_r}$ is the unit vector of the vector distance $\vec{r}$. The magnetic field at a specific point is proportional to the magnitude of the current and the length of the current element. The magnetic field is inversely proportional to the square of the distance of the point from the current element. The magnetic field value depends on the orientation angle of a specific point with respect to the current element.

### D. The Magnetic Force Law Between Two Current Elements

The magnetic force law is an equation that quantifies the force of attraction or repulsion between two current-carrying elements. This force occurs because one current element generates a magnetic field as defined by the Biot-Savart law, while the other element experiences a magnetic force as a result [12]. The magnetic force is defined as in equation (2) [13], [14].

$$\overrightarrow{dF_{12}} = \frac{\mu}{4\pi}\frac{I_1 I_2}{|\overrightarrow{r}|^2}(\overrightarrow{dl_2} \times (\overrightarrow{dl_1} \times \overrightarrow{a_r})) \tag{2}$$

where $\overrightarrow{dF_{12}}$ is the force felt by current element 2 due to current element 1. $I_1$ and $I_2$ are the amounts of current running in current elements 1 and 2, respectively. $\overrightarrow{dl_1}$ and $\overrightarrow{dl_2}$ are infinitesimal vectors that specify the directions of propagation for the current running through current elements 1 and 2, respectively. $\overrightarrow{r}$ is the distance vector pointing from current element 1 toward current element 2. $|\overrightarrow{r}|$ is the distance between these current elements. Equation (2) indicates that the magnetic force is perpendicular to both the direction of the affected current element and the magnetic field. The magnitude of the force is proportional to the sine of the angle between the current direction in the affected current element and the magnetic field. This implies that the magnetic force is zero if the current propagates in a direction parallel to the magnetic field.

## III. Related Work

Two attempts to explain the magnetic force between two current elements as an electric force were found: one attempt uses the special relativity theory [15], [16], and the second attempt uses a retarded action [6].

In [15], Kampen investigated the force between two parallel current-wires in the rest frames of the ions and the electrons. By applying the Lorentz transformation, the force appears as purely magnetostatic in the ion frame, while the force appears as a combined magnetostatic and electrostatic in the electron frame. In his work, Kampen provided an analysis for the force exerted on a charged particle in the field of a current-carrying wire. The current-carrying wire is modeled as being formed from fixed positive ions and free electrons moving at drift speed $v_d$. In the rest frame of the ions, the wire is electrically neutral, i.e., the positive and negative charge densities must be equal; otherwise, there would be an extra electrostatic field that the electrons move to neutralize. In the rest frame of the electrons, the positive ions are moving at speed $v_d$. Let $\lambda$ be the positive ion density in the rest frame of the ions. Then, the relativistic Fitzgerald-Lorentz contraction increases the ions density in the rest frame of the electrons, as described in equation (3).

$$\frac{\lambda}{\sqrt{1-(v_d/c)^2}} \approx \lambda + \frac{\lambda v_d^2}{2c^2}. \tag{3}$$

where $c$ is the speed of light. Alternately, the electrons are at rest in this electron frame. Therefore, its density decreases by $\lambda v_d^2/2c^2$ in comparison to the electron density in the rest frame of the ions. Let $q$ be a charged particle moving parallel to the current carrying wire at the speed of the electron drift velocity $v_d$. In the rest frame of the ions, the wire applies a magnetostatic force on $q$. In the rest frame of the electrons, the wire applies an electrostatic force on $q$, as defined in equation (4).

$$F = \mu\gamma\frac{q\lambda v_d^2}{2\pi r} \tag{4}$$

where $F$ is the force applied on charge $q$ due to the existence of the current-carrying wire and $\gamma$ is the Lorentz factor. Lorentz factor is defined as $\gamma = \frac{1}{\sqrt{1-(\frac{v_d}{c})^2}}$. For $v_d \ll c$,$\gamma$ is approximated to 1, i.e., $\gamma \approx 1$. Then, the force applied on charge $q$ is evaluated to $F = \mu\frac{q\lambda v_d^2}{2\pi r}$. This force is purely electrical and is identical in magnitude to the purely magnetic force in the rest frame of the ions. Thus, special relativity theory suggests that the force applied on a charge moving parallel to a current-carrying wire is magnetic or electric depending on the frame of reference. There are two shortcomings of this work: (1) it does not explain the force between two parallel current-wires as purely electrostatic and (2) it does not apply to a charged particle moving in a direction perpendicular to the current-carrying wire. The charged particle is observed as moving in the same direction in both the rest frame of the positive ions and the rest frame of the electrons. Thus, the force applied on the charged particle is described as a purely magnetic force, and this is not desirable.

In [6], Johnson proposed that the magnetic force between two current elements occurs due to the inhomogeneous propagation of the electric field from different parts of continuously distributed moving charges in a conductor. This inhomogeneous propagation causes a net difference between the field from the moving electrons and the immobile ions in a conductor. A retarded action at a distance between the two elements occurs, i.e., a fundamental difference in the propagation time of the force between the electric field generated by the moving electron in a conductor and the electric field generated by the immobile positive ions. Using this explanation, Johnson applied Coulomb's Law to derive a law of force between two straight conductors carrying a current as shown in equation (5).

$$d^2F_{12} = \frac{(\mu_0 I_1 I_2 \cos\theta \cos\psi ds_1 ds_2)}{4\pi R_{12}^2}\overrightarrow{a_R}. \tag{5}$$

where $ds_1$ and $ds_2$ are infinitesimal lengths for the first and second conductors, respectively. $I_1$ and $I_2$ are the amounts of current for the first and second conductors. $R_{12}$ and $\overrightarrow{a_R}$ are the distance and the unit direction between the centers of two current elements on the conductor, respectively. $\theta$ and $\psi$ are the angles that $\overrightarrow{a_R}$ make with the first and the second current elements. The drawback of this approach is that equation (5) is not equal to the well-known magnetic force law between two straight conductors carrying a current as stated by Johnson himself. Therefore, this approach is not used to explain the magnetic force between two current elements as an electric force.

Current literature on the field of electromagnetism[17], [4], [18] still considers the magnetic force different in origin from the electrostatic force between charges because the current-carrying wires are intrinsically charge-neutral. For example, in [17], Kiani analyzed the axial buckling behavior of doubly parallel current-carrying nano-wires in the presence of a longitudinal magnetic field. He obtained a formula for the magnetic forces on each nano-wire resulting from the transverse vibration of its neighboring nano-wire and the longitudinally exerted magnetic fields. The derivation of this formula treated magnetism and electricity as different aspects of the same subject.

To conclude, there is no existing work that explains the electric origin of the magnetic force in a way that facilitates obtaining the well-known magnetic force law. This paper provides two new theories and a new current representation to explain the magnetic force between two filamentary current elements as a result of electric force interactions between current charges. This explanation is proved by obtaining the magnetic force law using the basis of electric forces.

## IV. Methodology

This section describes a new approach to explain the magnetic force between two filamentary current elements as an electric force. This approach depends on three factors: (1) representing the steady current in a current element as a flow of equal positive and negative charges moving at the speed of light, (2) considering the relative speed between current charges, and (3) considering the pattern of the electric field spreading through the space generated by the moving positive and negative charges in a current element. The approach is proved by deriving the magnetic force law between two current elements and the Biot-Savart law using the electric force concept. This approach is helpful in unifying the concepts of electricity and magnetism.

### *A. Light-Speed Current Representation*

Light-speed current representation expresses a steady current as equal positive and negative charges moving in opposite directions at the speed of light inside a filamentary current element. Each charge is considered a point charge, does not have a finite size, and is massless. The point charge is assumed to be massless because according to Einstein's special theory of relativity, massless particles can only travel at the speed of light [19]. The speed of the charges is constant at all times. The filamentary current element is electrically neutral, i.e., has a zero net charge, all the time because the number and amount of the positive charges are equal to the number and amount of the negative charges inside the space, i.e., region, of the current element. This representation is able to model a current of any amount or form. This representation is important because it is able to equivalently represent any current by the minimum equal amount of positive and negative charges moving in opposite directions that is needed to produce the current. This minimum amount is obtained when these charges are moving at the maximum possible speed, which is at the speed of light.

**Definition**: Light-speed current representation characterizes a current as equal amounts of positive and negative massless point charges moving at the speed of light in a medium parallel to the current propagation direction through an open surface area, and its normal is in the direction of the current propagation, as in equation (6).

$$I = \frac{(+Q)}{2}\, ds\,(c) + \frac{(-Q)}{2}\, ds\,(-c). \tag{6}$$

where $c$ represents the speed of light, $ds$ represents the infinitesimal open surface area of the filamentary current element, and $Q$ represents the amount of carrier charges that move at the speed of light per unit volume.

Equation (6) provides an equivalent representation of the current using the minimum amount of charge needed to produce that current. This representation is consistent with the represented current properties, i.e., amount, zero net charge, and propagation speed. A unit of light-speed current is defined as the current produced by a positive one-half Coulomb charge moving at the speed of light in the direction of the current propagation and a negative one-half Coulomb in the opposite direction of the current propagation through a unit open surface area with its normal in the direction of the current propagation; refer to equation (7).

$$I_{sh} = \frac{(+1)}{2}(c) + \frac{(-1)}{2}(-c) \tag{7}$$

where $I_{sh}$ is a unit light-speed current. The unit of light-speed current is called the shadid. Equation (7) indicates that one unit light-speed current is equal to one Ampere multiplied by the speed of light, i.e., $1\, shadid = c\, ampere$. Light-speed current representation is useful to model the steady current that flows in a current element with the minimum amount of charges that is needed to produce that current. In this work, the physical space of a current element is equivalently modeled by an empty space that is free of charges except for the ones that are crossing its space.

An electric current is defined as a flow of electric charges across a surface [20], [21]. An electric current is carried out by moving negative charges, positive charges, or both negative and positive charges. However, regardless of how a current is carried out, there are two facts about electric filamentary current elements [22]: (1) they are electrically neutral, i.e., the space net charge density remains zero in the current element, and (2) currents of the same amount and direction have the same electric and magnetic effect. For example, conductors are materials that, although electrically neutral, possess a large number of mobile charges. In the presence of an applied electric field, a coordinated movement of the charges occurs, negative charges move in the opposite direction of positive charges, and the space net charge density remains zero at any point in the material [1], [22]. Let $I$ denote the amount of an electric current. Then, the amount of an electric current through an open surface is represented as the net positive charge passing through the surface per unit time, as in equation (8) [23], [24].

$$I = N\,q\,ds\,(\overrightarrow{v}.\overrightarrow{a_n}) \tag{8}$$

where $N$ is the number of charge carriers per unit volume, $q$ is the charge of the carriers, $\overrightarrow{v}$ is the average (drift) velocity of the carriers, $\overrightarrow{a_n}$ is the surface normal, and $ds$ is the infinitesimal surface area of the filamentary current element. The direction of the current is along the normal of the surface area. $Nq$ represents the amount of carrier charge per unit volume, then equation (8) is rewritten as equation (9).

$$I = Q\,ds\,(\overrightarrow{v}.\overrightarrow{a_n}). \tag{9}$$

where $Q$ is the amount of carrier charge per unit volume. Equation (9) indicates that different values for $Q$ and $\overrightarrow{v}$ are able to produce the same current if their multiplication as in equation (9) produces the same value, i.e., $I = Q_1 ds\,(\overrightarrow{v_1}.\overrightarrow{a_n})$=$Q_2 ds\,(\overrightarrow{v_2}.\overrightarrow{a_n})$. For example, let us assume there are two current elements with the same surface area, i.e., $\overrightarrow{a_{n1}} = \overrightarrow{a_{n2}} = \overrightarrow{a_n}$. Let the currents in these two elements be produced by a flow of positive charges in the same direction but at different speeds, i.e., $\angle\overrightarrow{v_1} = \angle\overrightarrow{v_2}$ but $|\overrightarrow{v_1}| \neq |\overrightarrow{v_2}|$. Then, the charges that flow in the second current element to produce the same current as in the first element are described by equation (10).

$$Q_1 = Q_2 \frac{|\overrightarrow{v_2}|}{|\overrightarrow{v_1}|}. \tag{10}$$

Equation (10) indicates that the same amount of current passing through an open surface can be produced by two different amounts of charges traveling at two different speeds. Another example is that the same amount of current produced by a flow of positive charges can be produced by a flow of negative charges traveling at the same speed but in the opposite direction, i.e., $Q_1 = -Q_2$ and $\overrightarrow{v_1} = -\overrightarrow{v_2}$.

In this work, a current is equally represented by one that is carried out by the minimum equal amount of positive and negative charges moving in opposite directions that is needed to produce that current. This minimum amount is obtained when these charges are moving at the maximum possible speed according to equation (8). The maximum possible speed is the speed of light [25]. This equivalent representation is useful in this work for two reasons: (1) it is consistent with the fact that current elements are electrically neutral because of the use of equal amount of positive and negative charges, and (2) the speed of charge is consistent with the speed of current propagation. Current propagation depends on the speed of the electric field wave that triggers, i.e., signals, the charge movement at each current point in the space to generate the same electric current, and the speed of this wave is the speed of light [26], [27], [28]. Thus, this representation is used throughout this work.

Light-speed current representation is able to equivalently model any current, even those generated by a moving charged particle. A current generated by a moving charged particle is modeled as being formed by a moving positive charge and a moving negative charge in opposite directions at the speed of light. Let $q$ be the measured charge of a moving particle at speed $v$. Let this particle move from position $\overrightarrow{p_1}$ to position $\overrightarrow{p_2}$, which is at infinitesimal distance $dl$ from $\overrightarrow{p_1}$, in the space during an infinitesimal time interval $dt$. At time $t$, the charged particle is at position $\overrightarrow{p_1}$ and there is no charge at position $\overrightarrow{p_2}$, while at time $t + dt$, the charged particle is at position $\overrightarrow{p_2}$ and there is no charge at position $\overrightarrow{p_1}$; see figure (3 a and b). Then, the movement of this charged particle is modeled by four charges: two static charges and two moving charges. The two static charges have a charge of $q/2$, one at the space of point $\overrightarrow{p_1}$ and the other one at the space of point $\overrightarrow{p_2}$. The two moving charges are opposite charges that flow between positions $\overrightarrow{p_1}$ and $\overrightarrow{p_2}$. At time $t$, a charge of $q/2$ is at the space of point $\overrightarrow{p_1}$, and a charge of $-q/2$ is at the space of point $\overrightarrow{p_2}$. Therefore, position $\overrightarrow{p_1}$ has a total electric charge of $q$, while position $\overrightarrow{p_2}$ is electrically neutral. At time $t + dt$, a charge of $q/2$ is at the space of point $\overrightarrow{p_2}$, and a charge of $-q/2$ is at the space of point $\overrightarrow{p_1}$. Therefore, position $\overrightarrow{p_2}$ has a total electric charge of $q$, while position $\overrightarrow{p_1}$ is electrically neutral; see figure (3 c and d). The movement of positive and negative charges between $\overrightarrow{p_1}$ and $\overrightarrow{p_2}$ is represented by a flow of charged current carriers, i.e., small charged packets of absolute amount $\left|\frac{qv}{2c}\right|$ move at the speed of light between $\overrightarrow{p_1}$ and $\overrightarrow{p_2}$ during the time interval $dt$ to move the charge $q/2$ from $\vec{p1}$ to $\vec{p2}$ and the charge $-q/2$ from $\vec{p2}$ to $\vec{p1}$; see figure (4). This model represents a smooth movement of the charged particle from $\vec{p1}$ to $\vec{p2}$ using positive and negative charges moving at the speed of light.

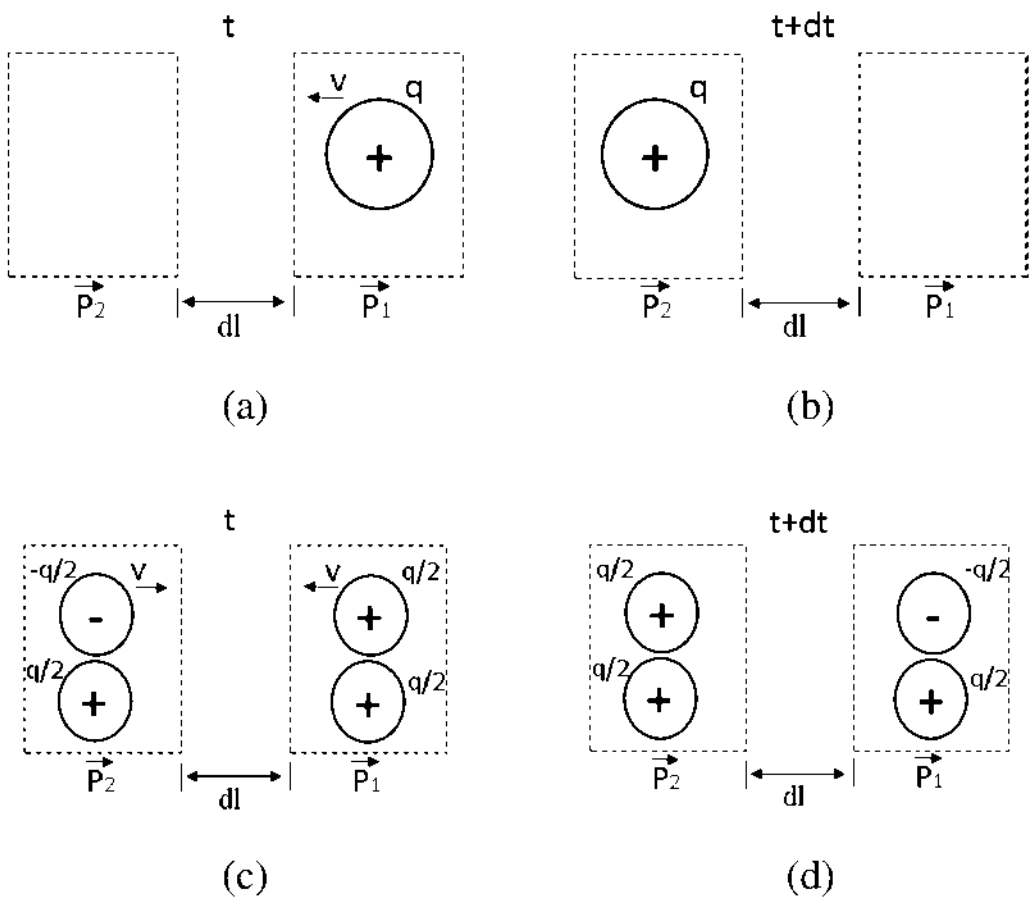

Figure 3: Shows the model that represents a moving charged particle as the movement of positive and negative charges. (a) shows the charged particle at position $\overrightarrow{p_1}$ moving toward the empty position $\overrightarrow{p_2}$ at infinitesimal distance $dl$ at time $t$. (b) shows the charged particle when arriving at position $\overrightarrow{p_2}$ after the infinitesimal time interval $dt$. (c) represents the movement of the charged particle at time $t$ by two static charges of amount $q/2$ in the vicinity of both positions and two moving charges, one of amount $q/2$ in the vicinity of position $\overrightarrow{p_1}$ and moving toward $\overrightarrow{p_2}$ and the other one of amount $-q/2$ in the vicinity of position $\overrightarrow{p_2}$ and moving toward $\overrightarrow{p_1}$. The position $\overrightarrow{p_1}$ has a charge of $q$, while the position $\overrightarrow{p_2}$ is electrically neutral. (d) represents the situation of the charges after the infinitesimal time interval $dt$; the moving charge of $-q/2$ arrives at position $\overrightarrow{p_1}$, while the moving charge of $q/2$ arrives at position $\overrightarrow{p_2}$. The position $\overrightarrow{p_2}$ has a charge of $q$, while position $\overrightarrow{p_1}$ is electrically neutral.

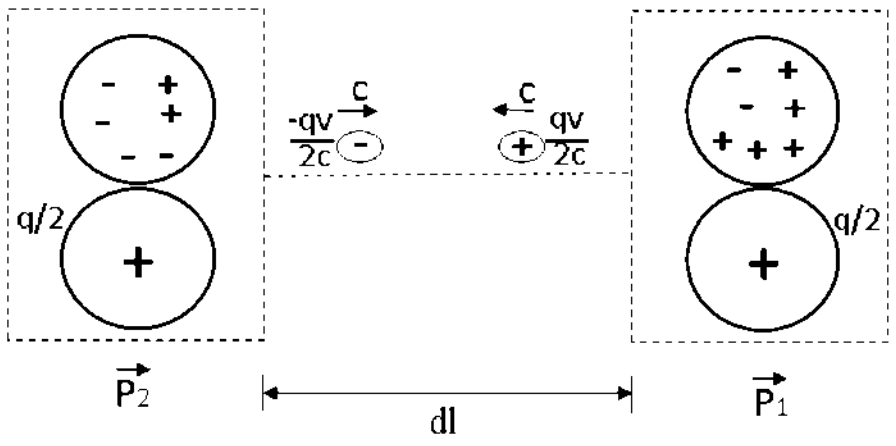

Figure 4: Shows an illustration of the flow of positive and negative charges at the speed of light to move a charge of amount $q/2$ from $\vec{p1}$ to $\vec{p2}$ and to move a charge of amount $-q/2$ from $\vec{p2}$ to $\vec{p1}$. $\vec{p1}$ and $\vec{p2}$ are an infinitesimal distance apart. This movement produces the same exact current generated by the movement of a charged particle $q$ at speed $v$ from $\vec{p1}$ to $\vec{p2}$ during the infinitesimal time interval $dt$.

### B. Current Charge Relativity

Current charge relativity is a theory that specifies the net charge of a light-speed current in the rest frame of an observer. The theory determines the relative net charge for a current element based on the number of crossing charges for the space of that element in the rest frame for an observer. The current has a zero net charge with respect to a static observer, to an observer moving in a perpendicular direction to the observed current propagation, or to an observer moving in a parallel direction to the observed current propagation at a speed less than the speed of light, a negative net charge with respect to an observer moving at the speed of light in the direction of positive charges of the current, and a positive net charge with respect to an observer moving at the speed of light in the direction of negative charges of the current. Light-speed current is composed of the same amounts of positive and negative charges moving in opposite directions at the speed of light crossing the space of a current element; see section (IV-A). Current charge relativity theory takes into consideration the velocity of the charges that form a light-speed current and the velocity of an observer for that current. The current charge relativity theory states the following:

**Theory (1)**: Light-speed current has a relative net charge with respect to the motion of an observer based on the number of crossing charges for the space of a current element in the rest frame for that observer. There are three cases: (1) the current has a zero net charge with respect to a static observer, an observer moving in a perpendicular direction to the observed current propagation, or to an observer moving in a parallel direction to the observed current propagation at a speed less than the speed of light, (2) the current has a negative net charge with respect to an observer moving at the speed of light in the direction of the current propagation, and (3) the current has a positive net charge with respect to an observer moving at the speed of light in the opposite direction of the current propagation.

**Proof**: Let $I$ represent the observed light-speed current and be defined by equation (11).

$$I = Q\,ds\,I_{sh} = \frac{(+Q)}{2}\,ds\,(c) + \frac{(-Q)}{2}\,ds\,(-c) \qquad (11)$$

Let $S_p$ be the speed of an observer. Then, there are six possible situations for the motion of a current observer: (1) a static observer, (2) an observer moving in a perpendicular direction to the observed current propagation, (3) an observer moving in the direction of the current propagation at a speed less than the speed of light, (4) an observer moving in the opposite direction of the current propagation at a speed less than the speed of light, (5) an observer moving at the speed of light in the direction of the current propagation, and (6) an observer moving at the speed of light in the opposite direction of the current propagation. The current charge that is observed in the relative frame of the observer is computed for each of these motion situations.

In the first motion situation, for a static observer, $S_p = 0$ along the direction of the current propagation, and then,

$$I = \frac{(+Q)}{2}\,ds\,\frac{(c-0)}{1-\frac{c\,0}{c^2}} + \frac{(-Q)}{2}\,ds\,\frac{(-c-0)}{1-\frac{c\,0}{c^2}}$$

$$I = \frac{(+Q)}{2}\,ds\,(c) + \frac{(-Q)}{2}\,ds\,(-c)$$

Then, the net charge per volume, denoted as $NCV$, for the current that is observed in the relative frame of the static observer is,

$$NCV = \frac{(+Q)}{2} + \frac{(-Q)}{2} = 0$$

In the second motion situation, for an observer moving in a perpendicular direction to the observed current propagation with speed $S_p$,

$$I = \frac{(+Q_p)}{2}\,ds\,(c\sqrt{1-(\frac{S_p}{c})^2}) + \frac{(-Q_p)}{2}\,ds\,(-c\sqrt{1-(\frac{S_p}{c})^2})$$

Where $Q_p$ is the charge density in the relative frame of the moving observer. Charge speeds in this frame are computed according to the theory of relativity [29], [30]. Since the current value is the same in all relative frames, $Q_p$ is rewritten in terms of the charge density $Q$ that is defined by equation (11) as follows:

$$Q_p\,ds\,c\sqrt{1-(\frac{S_p}{c})^2} = Q\,ds\,c$$

$$Q_p = \frac{Q}{\sqrt{1-(\frac{S_p}{c})^2}}$$

Then, for an observer moving in a perpendicular direction to the observed current propagation with speed $S_p$,

$$I = \frac{(+Q)}{2\sqrt{1-(\frac{S_p}{c})^2}}\,ds\,(c\sqrt{1-(\frac{S_p}{c})^2}) + \frac{(-Q)}{2\sqrt{1-(\frac{S_p}{c})^2}}\,ds\,(-c\sqrt{1-(\frac{S_p}{c})^2})$$

Then, the net charge per volume for the current that is observed in the relative frame of this moving observer is,

$$NCV = \frac{(+Q)}{2\sqrt{1-(\frac{S_p}{c})^2}} + \frac{(-Q)}{2\sqrt{1-(\frac{S_p}{c})^2}} = 0$$

In the third motion situation, for an observer moving in the direction of the current propagation with speed $S_p < c$ ,

$$I = \frac{(+Q_p)}{2}\, ds\,(\frac{c - S_p}{1-\frac{c\,S_p}{c^2}}) + \frac{(-Q_p)}{2}\, ds\,(-\frac{(c+S_p)}{1+\frac{c\,S_p}{c^2}})$$

$$I = \frac{(+Q_p)}{2}\, ds\,(c) + \frac{(-Q_p)}{2}\, ds\,(-c)$$

Then, $Q_p$ is rewritten in terms of the charge density $Q$ that is defined by equation (11) as follows:

$$Q_p\, ds\, c = Q\, ds\, c$$

$$Q_p = Q$$

Then, the net charge per volume for the current that is observed in the relative frame of this moving observer is,

$$NCV = \frac{(+Q)}{2} + \frac{(-Q)}{2} = 0$$

In the fourth motion situation, for an observer moving in the opposite direction of the current propagation with speed $S_p < c$ ,

$$I = \frac{(+Q_p)}{2}\, ds\,(\frac{c + S_p}{1+\frac{c\,S_p}{c^2}}) + \frac{(-Q_p)}{2}\, ds\,(-\frac{(c-S_p)}{1-\frac{c\,S_p}{c^2}})$$

$$I = \frac{(+Q_p)}{2}\, ds\,(c) + \frac{(-Q_p)}{2}\, ds\,(-c)$$

Then, $Q_p$ is rewritten in terms of the charge density $Q$ that is defined by equation (11) as follows:

$$Q_p\, ds\, c = Q\, ds\, c$$

$$Q_p = Q$$

Then, the net charge per volume for the current that is observed in the relative frame of this moving observer is,

$$NCV = \frac{(+Q)}{2} + \frac{(-Q)}{2} = 0$$

In the fifth motion situation, for an observer moving in the direction of the current propagation with speed $S_p = c$ ,

$$I = NaN + \frac{(-Q_p)}{2}\, ds\,(-\frac{(c+c)}{1+\frac{c\,c}{c^2}})$$

$$I = \frac{(-Q_p)}{2}\, ds\,(-c)$$

Where $NaN$ refers to an undefined or unrepresentable value. The positive charges of the current element are not seen in the relative frame of an observer moving at the speed of light in the direction of the current propagation. That is because the relative speed of the positive charges with respect to this moving observer is not defined, i.e., $\frac{c-c}{1-\frac{c\,c}{c^2}} = \frac{0}{0} = NaN$. Therefore, as there is no speed, it does not exist [31], [30]. Notice that, from any other point of view, the difference in velocity, and thus speed, between the positive charges and the observer that are defined to be traveling in the same direction is zero.

Then, $Q_p$ is rewritten in terms of the charge density $Q$ that is defined by equation (11) as follows:

$$\frac{Q_p}{2}\, ds\, c = Q\, ds\, c$$

$$Q_p = 2\,Q$$

Then, the net charge per volume for the current that is observed in the relative frame of this moving observer is,

$$NCV = \frac{(-Q_p)}{2} = -Q$$

In the sixth motion situation, for an observer moving in the opposite direction of the current propagation with speed $S_p = c$ ,

$$I = \frac{(+Q_p)}{2}\, ds\,(\frac{(c+c)}{1+\frac{c\,c}{c^2}}) + NaN$$

$$I = \frac{(+Q_p)}{2}\, ds\,(c)$$

The negative charges of the current element are not seen in the relative frame of an observer moving at the speed of light in the opposite direction of the current propagation. That is because the relative speed of the negative charges with respect to the moving observer is not defined, i.e., $\frac{c-c}{1-\frac{c\,c}{c^2}} = \frac{0}{0} = NaN$.

Then, $Q_p$ is rewritten in terms of the charge density $Q$ that is defined by equation (11) as follows:

$$\frac{Q_p}{2}\, ds\, c = Q\, ds\, c$$

$$Q_p = 2\,Q$$

Then, the net charge per volume for the current that is observed in the relative frame of this moving observer is,

$$NCV = \frac{(+Q_p)}{2} = +Q$$

**Done.**

In theory (1), a current is modeled by an equal number of positive and negative charges carrying the same amount of charge and crossing the space of a current element at the speed of light but in opposite directions. Current propagation is decided by the direction of movement of the positive charges. There are six situations analyzed for the motion of a current observer: (1) a static observer, (2) an observer moving in a perpendicular direction to the observed current propagation, (3) an observer moving in the direction of the current propagation at a speed less than the speed of light, (4) an observer moving in the opposite direction of the current propagation at a speed less than the speed of light, (5) an observer moving at the speed of light in the direction of the current propagation, and (6) an observer moving at the speed of light in the opposite direction of the current propagation. In the first situation, if a static observer **p** observes the current charges from an observation point $(x, y, z)$ in the space, the relative speed of the positive charges with respect to $\mathbf{p}$ is $c$ in the direction of the current, while the relative speed of the negative charges with respect to $\mathbf{p}$ is $c$ in the opposite direction of the current. Then, a static observer $\mathbf{p}$ sees the same number of positive and negative charges crossing the space of the current element. Given that, a static observer sees a zero net charge for the current element. In the second situation, for an observer moving in a perpendicular direction to the current propagation, i.e., the observer has zero speed along the line parallel to the direction of current propagation, this observer sees the same number of positive and negative charges crossing the space of the current element. Given that, the observer sees a zero net charge for the current element. In the third situation, if a moving observer $\mathbf{p}$ with a fixed speed $S_p$ that is less than the speed of light in the direction of the current propagation, i.e., the direction of the positive charge movement, the relative speed of the positive charges with respect to $\mathbf{p}$ is $c$ in the direction of the current according to the relativity theory, while the relative speed of the negative charges with respect to $\mathbf{p}$ is $c$ in the opposite direction of the current. Then, the observer sees the same number of positive and negative charges crossing the space of the current element. Given that, this observer sees a zero net charge for the current element. Similarly, in the fourth situation, if a moving observer $\mathbf{p}$ with a fixed speed $S_p$ that is less than the speed of light in the direction opposite that of the current propagation, i.e., the direction of the negative charge movement, the relative speed of the positive charges with respect to $\mathbf{p}$ is $c$ in the direction of the current according to the relativity theory, while the relative speed of the negative charges with respect to $\mathbf{p}$ is $c$ in the opposite direction of the current. Then, the observer sees the same number of positive and negative charges crossing the space of the current element. Given that, this observer sees a zero net charge for the current element. In the fifth situation, if a moving observer $\mathbf{p}$ with the speed of light, i.e., $S_p = c$, in the direction of the current propagation, i.e., the direction of the positive charge movement, the relative speed of the positive charges with respect to $\mathbf{p}$ is not defined, i.e., there is no speed for the positive charges, while the relative speed of the negative charges with respect to $\mathbf{p}$ is $c$ in the opposite direction of the current. Then, the observer sees only negative charges crossing the space of the current element. Given that, this observer sees a negative net charge for the current element. Notice that the density of the current charges is doubled in the relative frame for this moving observer because the current value is the same in all relative frames. In the sixth situation, if a moving observer $\mathbf{p}$ with the speed of light, i.e., $S_p = c$, in the direction opposite that of the current propagation, i.e., the direction of the negative charge movement, the relative speed of the positive charges with respect to $\mathbf{p}$ is $c$ in the direction of the current, while the relative speed of the negative charges with respect to $\mathbf{p}$ is not defined, i.e., there is no speed for the negative charges. Then, the observer sees only positive charges crossing the space of the current element. Given that, this observer sees a positive net charge for the current element. Notice that the density of the current charges is doubled in the relative frame for this moving observer because the current value is the same in all relative frames. As a result, a current element is modeled as being formed from the same number of positive and negative charges crossing the space of the current element with respect to a static observer, to an observer moving in a perpendicular direction to the observed current propagation, or to an observer moving in a parallel direction to the observed current propagation at a speed less than the speed of light. A current element is modeled as being completely formed from negative charges with respect to an observer moving at the speed of light in the direction of the current propagation, while a current element is modeled as being completely formed from positive charges with respect to an observer moving at the speed of light in the opposite direction of the current propagation. Figure (5) shows the observed charges seen by an observer $\mathbf{p}$ in the six motion situations.

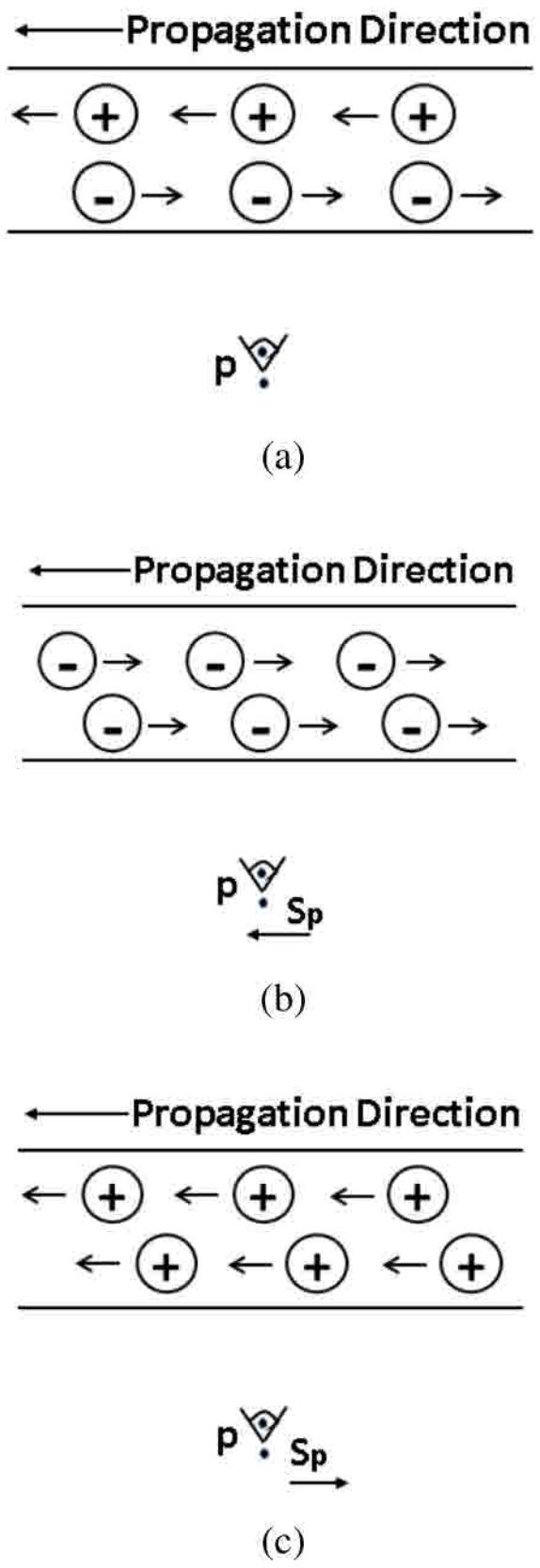

Figure 5: Shows the relative current charges seen by observation point $p$ moving at different speeds $S_p$. (a) Relative current charges seen with respect to a static observer, to an observer moving in a perpendicular direction to the observed current propagation, or to an observer moving in a parallel direction to the observed current propagation at a speed less than the speed of light. (b) Relative current charges seen at $S_p = c$ in the direction of the current propagation. (c) Relative current charges seen at $S_p = c$ in the opposite direction of the current propagation.

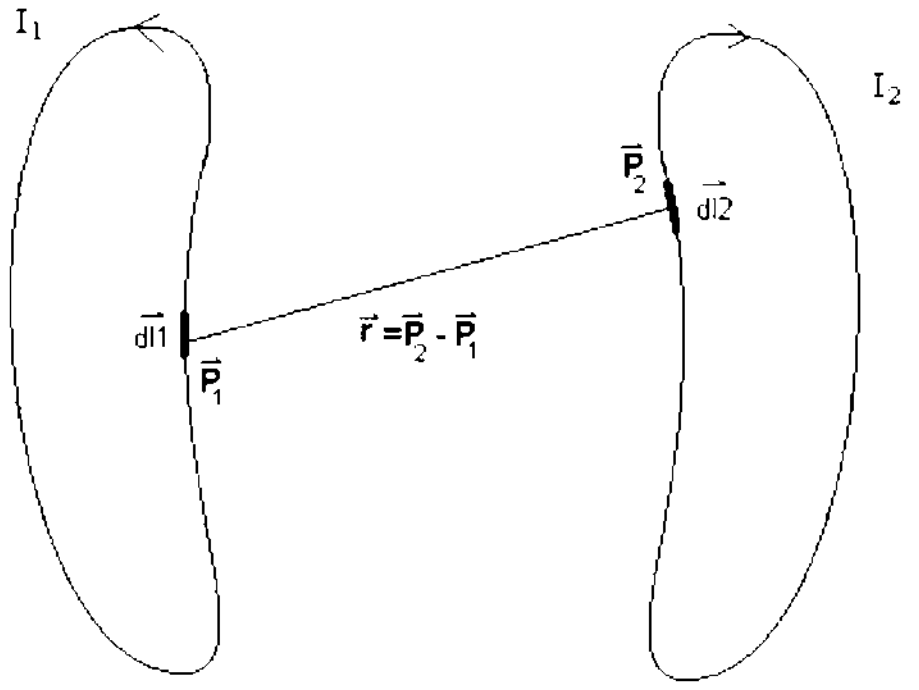

Figure 6: Shows two closed loops of constant filamentary current and two arbitrarily oriented current elements.

### *C. The Electric Origin of Magnetic Force*

This section provides a proof of the electric origin of magnetic force by deriving the Biot-Savart law using the electric force concept. This derivation is used to prove that the magnetic force is a result of interactions between electric forces. The derivation process of the Biot-Savart law analyzes the electric force between the charges of two filamentary current elements at different basic positions. Then, the region is found in which the electric force is observed for a current element to derive the Biot-Savart law.

Mathematical notations are provided for filamentary current elements, infinitesimal volume, and infinitesimal charge crossing the space of a current element to simplify the derivation process. Let $\overrightarrow{dI_1}$ and $\overrightarrow{dI_2}$ represent two filamentary current elements in a free space at positions $\overrightarrow{p_1}$ and $\overrightarrow{p_2}$, respectively, with a displacement vector $\overrightarrow{r}$ from $\overrightarrow{dI_1}$ to $\overrightarrow{dI_2}$, i.e., $\overrightarrow{r} = \overrightarrow{p_2} - \overrightarrow{p_1}$; refer to figure (6). The current elements $\overrightarrow{dI_1}$ and $\overrightarrow{dI_2}$ are defined in equations (12 and 13), respectively.

$$\overrightarrow{dI_1} = I_1\, dl\, \overrightarrow{a_1} \tag{12}$$

$$\overrightarrow{dI_2} = I_2\, dl\, \overrightarrow{a_2} \tag{13}$$

where $I_1$ and $I_2$ are the amounts of current for $\overrightarrow{dI_1}$ and $\overrightarrow{dI_2}$ , respectively. $dl$ is the infinitesimal length for the current elements $\overrightarrow{dI_1}$ and $\overrightarrow{dI_2}$. $\overrightarrow{a_1}$ and $\overrightarrow{a_2}$ are unit vectors for the directions of the current propagation for $\overrightarrow{dI_1}$ and $\overrightarrow{dI_2}$ , respectively. The amounts of the currents $I_1$ and $I_2$ are defined using the light-speed current representation, as shown in equations (14 and 15).

$$I_1 = \frac{Q_1}{2}\, ds\, c + \frac{(-Q_1)}{2}\, ds\, c. \tag{14}$$

$$I_2 = \frac{Q_2}{2}\, ds\, c + \frac{(-Q_2)}{2}\, ds\, c. \tag{15}$$

where $Q_1$ and $Q_2$ represent the amounts of charges for currents $I_1$ and $I_2$ that move at the speed of light per unit volume.

The infinitesimal volume for the current elements is defined in equation (16).

$$dV = ds\, dl. \tag{16}$$

where $dV$ is the infinitesimal volume. The infinitesimal volume of a current element is modeled as an empty region when there is no current, i.e., when it does not contain any charge, and this region is crossed by charges when a current exists.

The total amounts of charges that move at the speed of light contained in the infinitesimal volume for $\overrightarrow{dI_1}$ and $\overrightarrow{dI_2}$ are defined in equations (17 and 18).

$$dQ_1 = Q_1\, ds\, dl. \tag{17}$$

$$dQ_2 = Q_2\, ds\, dl. \tag{18}$$

where $dQ_1$ and $dQ_2$ are the total infinitesimal amounts of charge contained in the infinitesimal volumes for $\overrightarrow{dI_1}$ and $\overrightarrow{dI_2}$, respectively. An infinitesimal charge crosses the volume of a current element only when the current exists.

The derivation process analyzes the force of $\overrightarrow{dI_1}$ on $\overrightarrow{dI_2}$ using a 3D space formed from two perpendicular planes: (1) the horizontal plane and (2) the vertical plane. The horizontal plane fully contains the current element $\overrightarrow{dI_1}$ and the point $\overrightarrow{p_2}$. The normal vector for the horizontal plane is defined in equation (19).

$$\overrightarrow{a_n} = \frac{\overrightarrow{a_1} \times \overrightarrow{a_r}}{|\overrightarrow{a_1} \times \overrightarrow{a_r}|}. \tag{19}$$

where $\overrightarrow{a_n}$ is the normal for the horizontal plane. The vertical plane is perpendicular to the horizontal plane, and its normal is defined in equation (20).

$$\overrightarrow{a_{np}} = \frac{\overrightarrow{a_n} \times \overrightarrow{a_1}}{|\overrightarrow{a_n} \times \overrightarrow{a_1}|}. \tag{20}$$

where $\overrightarrow{a_{np}}$ is the normal for the vertical plane. The three orthonormal vectors that define the 3D space are: (1) $\overrightarrow{u_1} = \overrightarrow{a_1}$, (2) $\overrightarrow{u_2} = \overrightarrow{a_{np}}$, and (3) $\overrightarrow{u_3} = \overrightarrow{a_n}$. The current element $\overrightarrow{dI_2}$ is decomposed into its three perpendicular components in the 3D space: (1) the component along $\overrightarrow{u_1}$ that is parallel to $\overrightarrow{dI_1}$, referred to as the parallel component, (2) the component along $\overrightarrow{u_2}$ that is perpendicular to $\overrightarrow{dI_1}$ and lies on the horizontal plane, referred to as the horizontal perpendicular component, and (3) the component along $\overrightarrow{u_3}$ that is perpendicular to $\overrightarrow{dI_1}$ and lies on the vertical plane, referred to as the vertical perpendicular component. These planes and components are shown in figure (7). The derivation process derives the force law for each component for two cases: (1) $|\overrightarrow{a_1} \times \overrightarrow{a_r}| = 1$ and (2) $\overrightarrow{a_1} \times \overrightarrow{a_r} = 0$. Thus, there are six cases:

1) Two parallel filamentary current elements where $|\overrightarrow{a_1} \times \overrightarrow{a_r}| = 1$,
2) Two parallel filamentary current elements where $\overrightarrow{a_1} \times \overrightarrow{a_r} = 0$,
3) Two perpendicular filamentary current elements on one plane where $|\overrightarrow{a_1} \times \overrightarrow{a_r}| = 1$,
4) Two perpendicular filamentary current elements on one plane where $\overrightarrow{a_1} \times \overrightarrow{a_r} = 0$,

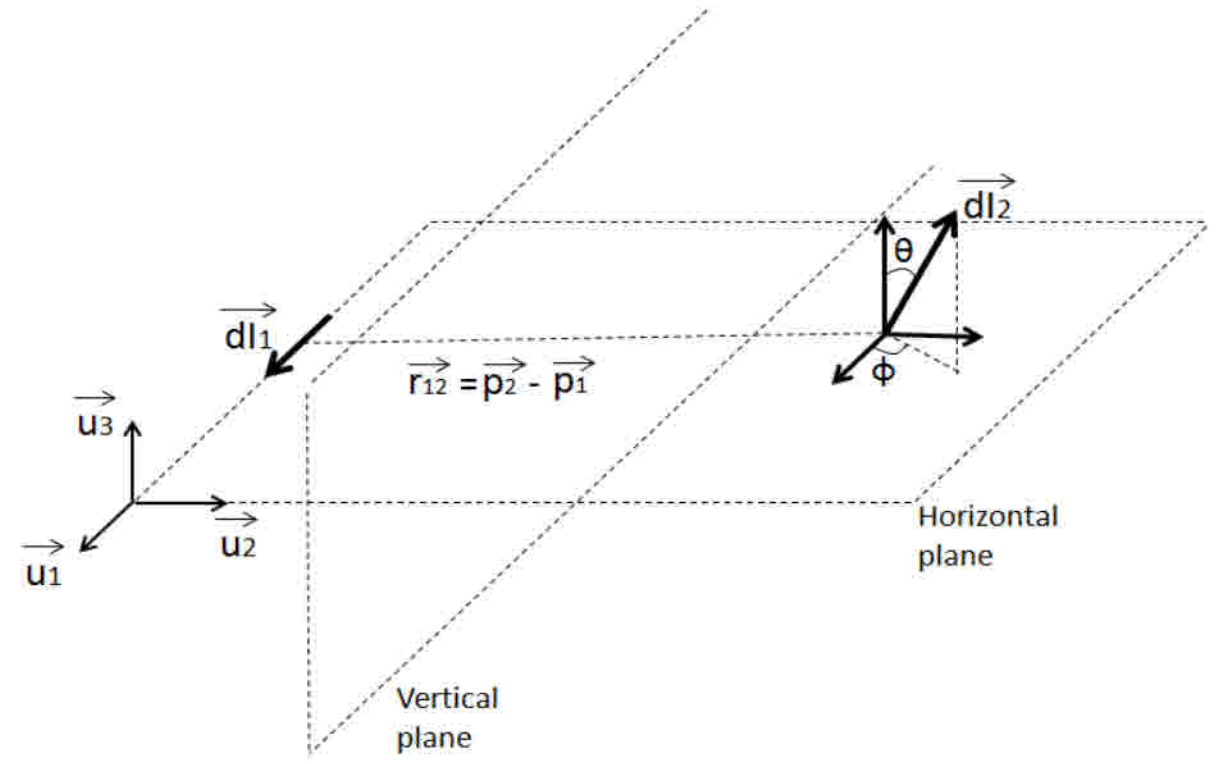

Figure 7: Shows the 3D space and the three perpendicular components for current elements.

5) Two perpendicular filamentary current elements on two perpendicular planes where $|\overrightarrow{a_1} \times \overrightarrow{a_r}| = 1$,
6) Two perpendicular filamentary current elements on two perpendicular planes where $\overrightarrow{a_1} \times \overrightarrow{a_r} = 0$.

These cases are shown in figure (8). Notice that case (6) is another case for two perpendicular filamentary current elements on one plane, similar to case (4). Case (6) is mentioned to show what happens for two perpendicular filamentary current elements on two perpendicular planes when $\overrightarrow{a_1} \times \overrightarrow{a_r} = 0$. The discussion of these cases is organized into three sections, followed by a section that derives the Biot-Savart law.

*1) Two Parallel Filamentary Current Elements:* This section derives the electric force law between two parallel filamentary current elements. The force law is derived for case (1) and case (2) of the six cases mentioned earlier, and then the general force law for two parallel elements is found.

For case (1), let $\overrightarrow{dI_1}$ and $\overrightarrow{dI_2}$ be two parallel current elements that are fully contained in one plane, and $\overrightarrow{dI_1}$ is perpendicular to $\overrightarrow{r}$ , i.e., $|\overrightarrow{a_1} \times \overrightarrow{a_r}| = 1$, as shown in figure (9 a). The electric force is computed for two situations for current propagation: (1) the currents propagate in the same direction, and (2) the currents propagate in opposite directions. For each situation, the electric charges are computed, and then the electric force between them is computed.

For the first situation, according to the current charge relativity theory mentioned in section (IV-B), when the charges of $\overrightarrow{dI_2}$ are observing and affected by the charges in current $\overrightarrow{dI_1}$, the moving charges of $\overrightarrow{dI_1}$ appear completely negative in the relative frame for the positive moving charges of $\overrightarrow{dI_2}$, as computed in equation (21); see figure (9 b).

$$I_{12+} = (-Q_1)\, ds\, (-c) \tag{21}$$

where $I_{12+}$ is the light-speed representation for the current of $\overrightarrow{dI_1}$ in the relative frame for the positive charges of $\overrightarrow{dI_2}$. Notice that the negative charge of $\overrightarrow{dI_2}$ is not included in the relative frame for the positive charge of $\overrightarrow{dI_2}$ because these charges are assumed not to be feeling the existence of each other. This can be explained in part by the following. These charges are moving at the speed of light in opposite directions, which is the same as the speed of spreading of the electric field in the space [32], [23]. These charges have an infinitesimal distance between them. Therefore, by the time the effect of the electric field for the negative charge arrives at the position of the positive charge, the positive charge has already moved by switching positions with the negative charge to create the motion that produces the current $\overrightarrow{dI_2}$, hence the exclusion of the negative charge of $\overrightarrow{dI_2}$ from the relative frame for the positive charge of $\overrightarrow{dI_2}$. The movement and positions of positive and negative charges of a light speed current inside a current element are described later in section (IV-C2).

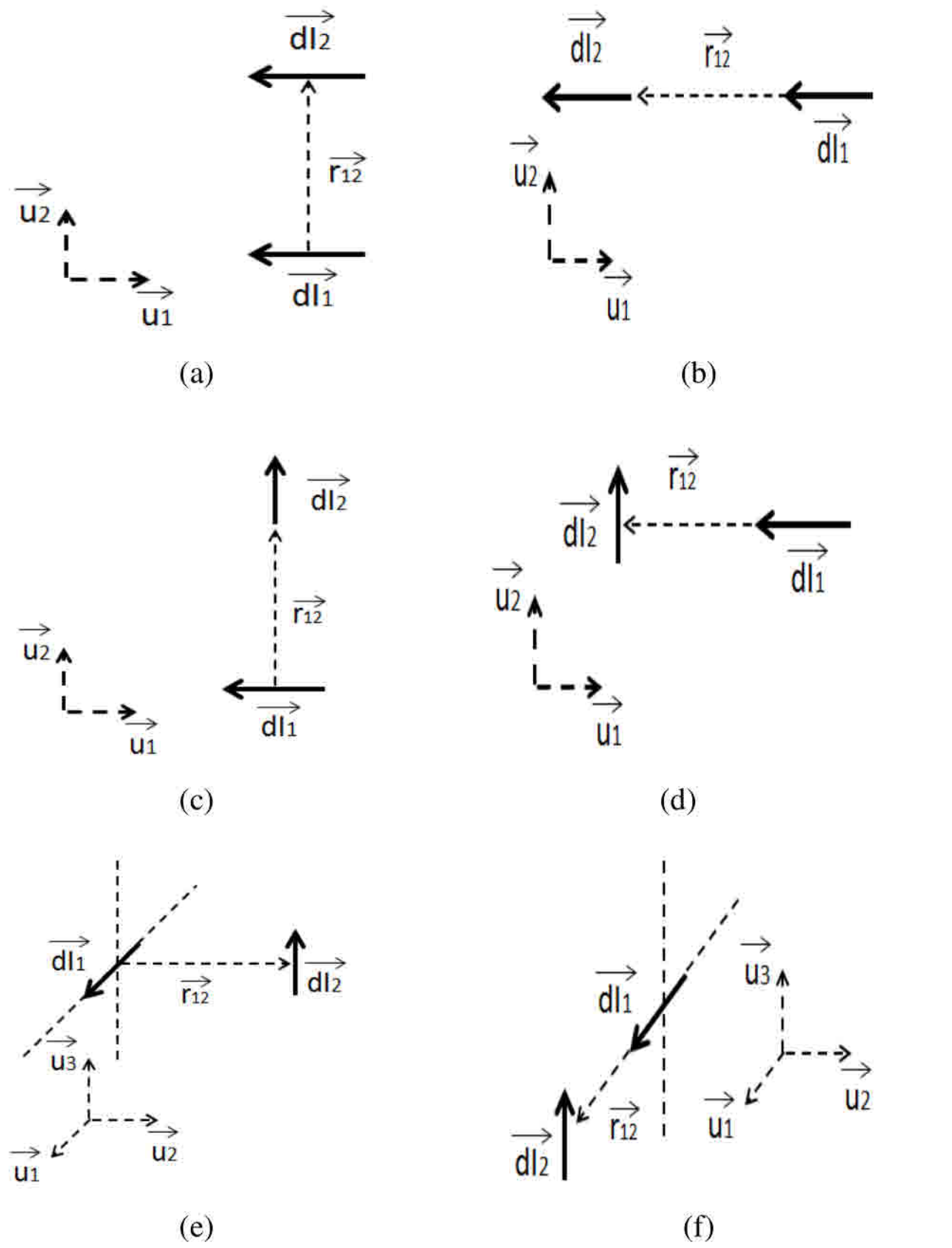

Figure 8: Shows the six basic relative position cases between two filamentary current elements $\overrightarrow{dI_1}$ and $\overrightarrow{dI_2}$. (a) Case-1 for two parallel filamentary current elements where $|\overrightarrow{a_1} \times \overrightarrow{a_r}| = 1$. (b) Case-2 for two parallel filamentary current elements where $\overrightarrow{a_1} \times \overrightarrow{a_r} = 0$. (c) Case-3 for two perpendicular filamentary current elements on one plane where $|\overrightarrow{a_1} \times \overrightarrow{a_r}| = 1$. (d) Case-4 for two perpendicular filamentary current elements on one plane where $\overrightarrow{a_1} \times \overrightarrow{a_r} = 0$. (e) Case-5 for two perpendicular filamentary current elements on two perpendicular planes where $|\overrightarrow{a_1} \times \overrightarrow{a_r}| = 1$. (f) Case-6 for two perpendicular filamentary current elements on two perpendicular planes where $\overrightarrow{a_1} \times \overrightarrow{a_r} = 0$.

The moving charges of $\overrightarrow{dI_1}$ appear completely positive in the relative frame for the negative moving charges of $\overrightarrow{dI_2}$, as computed in equation (22); see figure (9 c).

$$I_{12-} = Q_1\, ds\, c \tag{22}$$

where $I_{12-}$ is the light-speed representation for the current of $\overrightarrow{dI_1}$ in the relative frame for the negative charges of $\overrightarrow{dI_2}$. The infinitesimal relative charge of $\overrightarrow{dI_1}$ seen by the positive

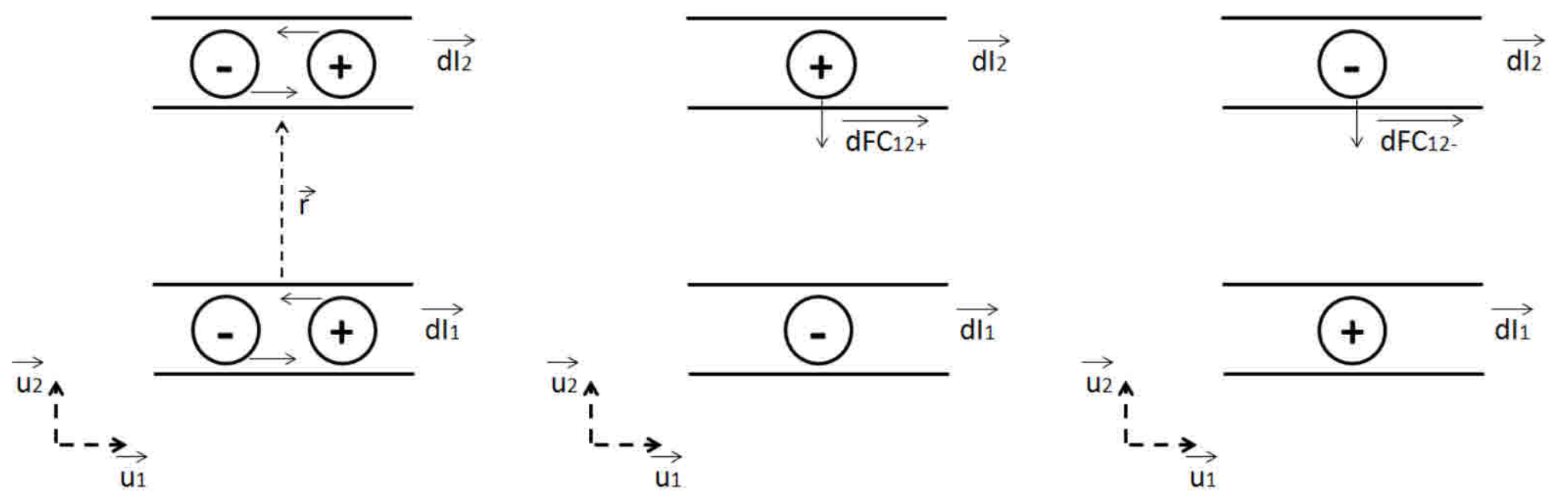

(a) (b) (c)

Figure 9: Shows case (1) for $\overrightarrow{dI_1}$ and $\overrightarrow{dI_2}$ with currents that propagate in the same direction. (a) Two parallel current elements $\overrightarrow{dI_1}$ and $\overrightarrow{dI_2}$ with currents that propagate on the negative $\overrightarrow{u_1}$-axis. (b) The relative negative current charge of $\overrightarrow{dI_1}$ seen by the positive charge of $\overrightarrow{dI_2}$. (c) The relative positive current charge of $\overrightarrow{dI_1}$ seen by the negative charge of $\overrightarrow{dI_2}$.

moving charges of $\overrightarrow{dI_2}$, denoted as $dQ_{12+}$, is computed as in equation (23).

$$dQ_{12+} = (-Q_1)\, ds\, dl \tag{23}$$

where $ds\, dl$ is the infinitesimal volume of the current element $\overrightarrow{dI_1}$. The infinitesimal positive charge of $\overrightarrow{dI_2}$, denoted as $dQ_{2+}$, is computed as in equation (24).

$$dQ_{2+} = (\frac{Q_2}{2})\, ds\, dl = \frac{dQ_2}{2} \tag{24}$$

where $dQ_2$ is the infinitesimal positive charge that is needed to generate the current $I_2$ in $\overrightarrow{dI_2}$.

The infinitesimal relative charge of $\overrightarrow{dI_1}$ seen by the negative moving charges of $\overrightarrow{dI_2}$, denoted as $dQ_{12-}$, is computed as in equation (25).

$$dQ_{12-} = (Q_1)\, ds\, dl \tag{25}$$

where $ds\, dl$ is the infinitesimal volume of the current element $\overrightarrow{dI_2}$. The two current elements have the same infinitesimal volume dimensions. The infinitesimal negative charge of $\overrightarrow{dI_2}$, denoted as $dQ_{2-}$, is computed as in equation (26).

$$dQ_{2-} = (\frac{-Q_2}{2})\, ds\, dl = -\frac{dQ_2}{2} \tag{26}$$

Then, $\overrightarrow{dI_1}$ has a negative relative charge with respect to the moving positive charges of $\overrightarrow{dI_2}$, and $\overrightarrow{dI_1}$ has a positive relative charge with respect to the moving negative charges of $\overrightarrow{dI_2}$. Thus, these charges attract each other.

The magnitude of the electric force between the relative charges of $\overrightarrow{dI_1}$, i.e., $dQ_{12+}$, and the moving positive charges of $\overrightarrow{dI_2}$, $dQ_{2+}$, is computed as in equation (27).

$$\left|\overrightarrow{dFC_{12+}}\right| = \frac{1}{4\,\pi\,\epsilon}\frac{|dQ_{12+}|\;|dQ_{2+}|}{|\overrightarrow{r}|^2},$$

$$\left|\overrightarrow{dFC_{12+}}\right| = \frac{1}{4\,\pi\,\epsilon}\frac{|-dQ_1|\;\left|\frac{dQ_2}{2}\right|}{|\overrightarrow{r}|^2},$$

$$\left|\overrightarrow{dFC_{12+}}\right| = \frac{1}{2}\frac{1}{4\,\pi\,\epsilon}\frac{dQ_1\, dQ_2}{|\overrightarrow{r}|^2}. \tag{27}$$

where $\left|\overrightarrow{dFC_{12+}}\right|$ is the magnitude of the electric force between $dQ_{12+}$ and $dQ_{2+}$, and $\epsilon$ is the electric permittivity. The electric force $\left|\overrightarrow{dFC_{12+}}\right|$ is an attraction force in the direction of the negative $\overrightarrow{u_2}$-axis because $dQ_{12+}$ is negative while $Q_{2+}$ is positive; refer to figure (9 b).

The magnitude of the electric force between the relative charges of $\overrightarrow{dI_1}$, i.e., $dQ_{12-}$, and the moving negative charges of $\overrightarrow{dI_2}$, $dQ_{2-}$, is computed as in equation (28).

$$\left|\overrightarrow{dFC_{12-}}\right| = \frac{1}{4\,\pi\,\epsilon}\frac{|dQ_{12-}|\;|dQ_{2-}|}{|\overrightarrow{r}|^2},$$

$$\left|\overrightarrow{dFC_{12-}}\right| = \frac{1}{4\,\pi\,\epsilon}\frac{|dQ_1|\;\left|-\frac{dQ_2}{2}\right|}{|\overrightarrow{r}|^2},$$

$$\left|\overrightarrow{dFC_{12-}}\right| = \frac{1}{2}\frac{1}{4\,\pi\,\epsilon}\frac{dQ_1\, dQ_2}{|\overrightarrow{r}|^2}. \tag{28}$$

where $\left|\overrightarrow{dFC_{12-}}\right|$ is the magnitude of the electric force between $dQ_{12-}$ and $dQ_{2-}$, and $\epsilon$ is the permittivity. The electric force $\left|\overrightarrow{dFC_{12-}}\right|$ is an attraction force in the direction of the negative $\overrightarrow{u_2}$-axis because $dQ_{12-}$ is positive, while $Q_{2-}$ is negative; refer to figure (9 c).

The electric forces $\left|\overrightarrow{dFC_{12+}}\right|$ and $\left|\overrightarrow{dFC_{12-}}\right|$ on the crossing charges of $\overrightarrow{dI_2}$ are perpendicular to their motion, and these charges are not permitted to leave their filamentary current element. Given that, these charges push the filamentary current element by these forces; refer to section (II-A).

The magnitude of the total electric force that is applied on the current element $\overrightarrow{dI_2}$ due to the existence of the current element $\overrightarrow{dI_1}$ is computed as shown in equation (29).

$$\left|\overrightarrow{dF_{12}}\right| = \left|\overrightarrow{dFC_{12+}}\right| + \left|\overrightarrow{dFC_{12-}}\right|,$$

$$\left|\overrightarrow{dF_{12}}\right| = \frac{1}{4\,\pi\,\epsilon}\frac{dQ_1\, dQ_2}{|\overrightarrow{r}|^2}. \tag{29}$$

where $\left|\overrightarrow{dF_{12}}\right|$ is the magnitude of the total electric force that is applied on $\overrightarrow{dI_2}$ by the perpendicular forces applied on its moving charges due to the existence of $\overrightarrow{dI_1}$. This force is in the direction of the negative $\overrightarrow{u_2}$-axis. To add an expression for the direction, equation (29) is rewritten and becomes equation (30).

$$\overrightarrow{dF_{12}} = -\frac{1}{4\,\pi\,\epsilon}\frac{dQ_1\,dQ_2}{|\overrightarrow{r}|^2}\,\overrightarrow{u_2}. \qquad (30)$$

where $\overrightarrow{dF_{12}}$ is the total attractive electric force, i.e., magnitude and direction.

The total electric force on $\overrightarrow{dI_2}$ due to $\overrightarrow{dI_1}$ is rewritten in terms of the current flowing through the current elements by multiplying and dividing by $c^2$ on the right side of equation (30), as shown in equation (31).

$$\overrightarrow{dF_{12}} = -\frac{c^2}{c^2}\frac{1}{4\,\pi\,\epsilon}\frac{dQ_1\,dQ_2}{|\overrightarrow{r}|^2}\,\overrightarrow{u_2}. \qquad (31)$$

Equation (31) is rewritten and simplified further as shown in equations (32 and 33).

$$\overrightarrow{dF_{12}} = -\frac{1}{4\,\pi\,(\epsilon\,c^2)}\frac{(Q_1\,ds\,c)\,(Q_2\,ds\,c)}{|\overrightarrow{r}|^2}\,dl\,dl\,\overrightarrow{u_2}. \qquad (32)$$

$$\overrightarrow{dF_{12}} = -\frac{1}{4\,\pi\,(\epsilon\,c^2)}\frac{I_1\,I_2}{|\overrightarrow{r}|^2}\,dl\,dl\,\overrightarrow{u_2}. \qquad (33)$$

Equation (33) is simplified further as shown in equation (34) because $\mu = \frac{1}{\epsilon\,c^2}$, where $\mu$ is the magnetic permeability.

$$\overrightarrow{dF_{12}} = -\frac{\mu}{4\,\pi}\frac{I_1\,I_2}{|\overrightarrow{r}|^2}\,dl\,dl\,\overrightarrow{u_2}. \qquad (34)$$

The electric force $\overrightarrow{dF_{12}}$ on $\overrightarrow{dI_2}$ due to $\overrightarrow{dI_1}$, as shown in equation (34), is equivalent in magnitude and direction to the magnetic force between two parallel infinitesimal current elements having currents that propagate in the same direction, as shown in figure (9 a).

For the second situation, by following a similar method as for the previous one, the electric force $\overrightarrow{dF_{12}}$ on $\overrightarrow{dI_2}$ due to $\overrightarrow{dI_1}$ is found when the currents of $\overrightarrow{dI_1}$ and $\overrightarrow{dI_2}$ propagate in opposite directions, as shown in figure (10 a). According to the current charge relativity theory mentioned in section (IV-B), the moving charges of $\overrightarrow{dI_1}$ appear completely positive in the relative frame for the positive moving charges of $\overrightarrow{dI_2}$; refer to figure (10 b). Meanwhile, the moving charges of $\overrightarrow{dI_1}$ appear completely negative in the relative frame for the negative moving charges of $\overrightarrow{dI_2}$; refer to figure (10 c). Thus, the relative charges of $\overrightarrow{dI_1}$ repel the charges of $\overrightarrow{dI_2}$, generating a repulsive force on the current element $\overrightarrow{dI_2}$. For the case shown in figure (10), this repulsive force is in the direction of the positive $\overrightarrow{u_2}$-axis. Then, the electric force $\overrightarrow{dF_{12}}$ on $\overrightarrow{dI_2}$ due to $\overrightarrow{dI_1}$ is computed via equation (35).

$$\overrightarrow{dF_{12}} = \frac{\mu}{4\,\pi}\frac{I_1\,I_2}{|\overrightarrow{r}|^2}\,dl\,dl\,\overrightarrow{u_2}. \qquad (35)$$

Equation (35) shows that the electric force $\overrightarrow{dF_{12}}$ on $\overrightarrow{dI_2}$ due to $\overrightarrow{dI_1}$ is equivalent in magnitude and direction to the magnetic force between two parallel infinitesimal current elements having currents that propagate in opposite directions, as shown in figure (10 a).

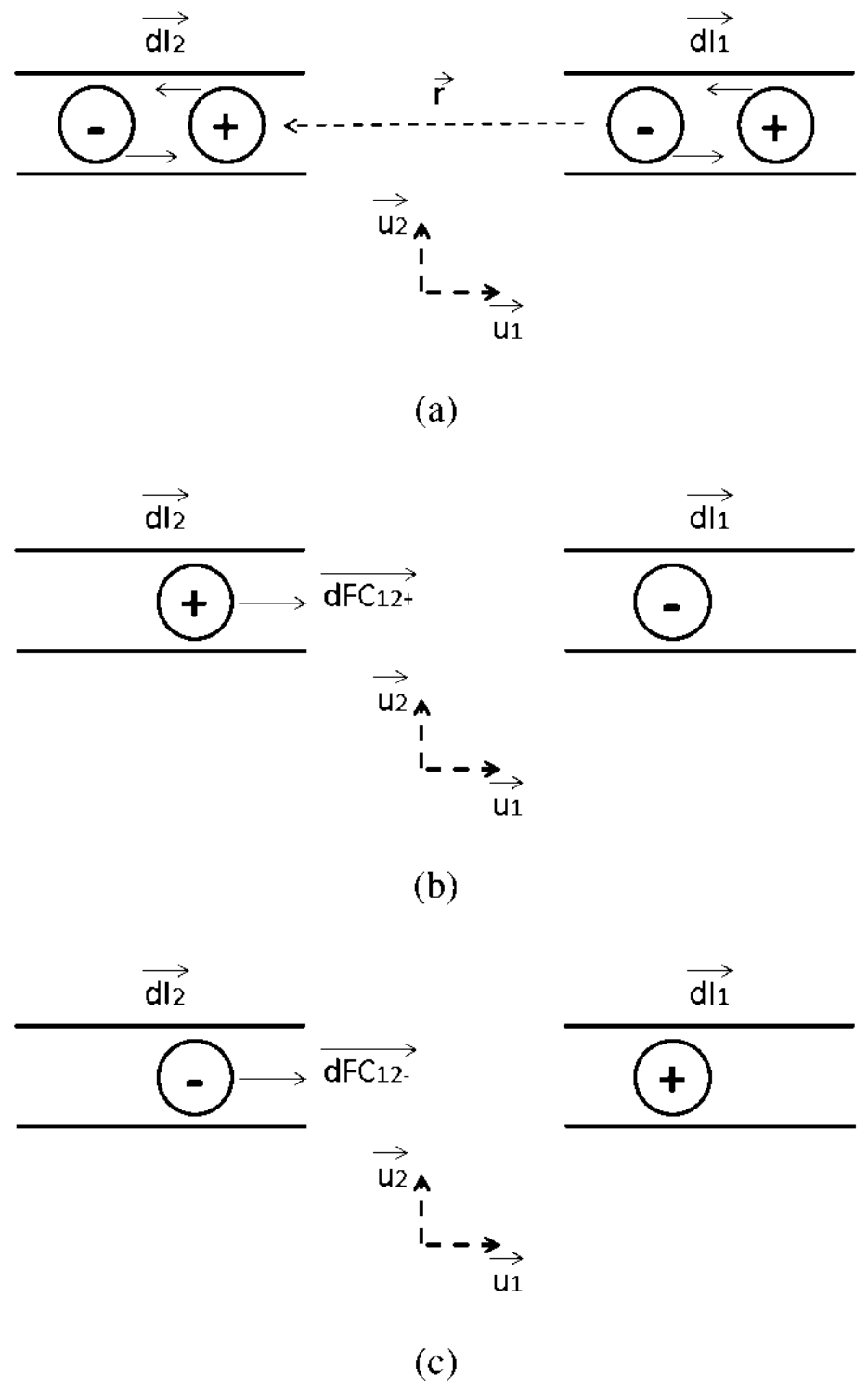

Figure 11: Shows case (2) for $\overrightarrow{dI_1}$ and $\overrightarrow{dI_2}$ with currents that propagate in the same direction. (a) Two parallel current elements $\overrightarrow{dI_1}$ and $\overrightarrow{dI_2}$ with currents that propagate along the negative $\overrightarrow{u_1}$-axis. (b) The relative negative current charge of $\overrightarrow{dI_1}$ seen by the positive charge of $\overrightarrow{dI_2}$. (c) The relative positive current charge of $\overrightarrow{dI_1}$ seen by the negative charge of $\overrightarrow{dI_2}$.

For case (2), $\overrightarrow{dI_2}$ and $\overrightarrow{dI_1}$ are parallel to each other, and $\overrightarrow{dI_1}$ is parallel to $\overrightarrow{r}$, i.e., $\overrightarrow{a_1} \times \overrightarrow{a_r} = 0$; see figure (11 a). The electric force on $\overrightarrow{dI_2}$ due to $\overrightarrow{dI_1}$ is computed for two situations for current propagation: (1) the currents propagate in the same direction, and (2) the currents propagate in opposite directions.

For the first situation, the analysis is similar to that performed for the first situation in case (1). According to the current charge relativity theory mentioned in section (IV-B), the moving charges of $\overrightarrow{dI_1}$ appear completely negative in the relative frame for the positive moving charges of $\overrightarrow{dI_2}$, as computed in equation (21); see figure (11 b). The moving charges of $\overrightarrow{dI_1}$ appear completely positive in the relative frame for the negative moving charges of $\overrightarrow{dI_2}$, as computed in equation (22); see figure (11 c). Then, these charges attract each other.

The magnitude of the electric force $\left|\overrightarrow{dFC_{12+}}\right|$ that the relative negative charge of $\overrightarrow{dI_1}$ affects the positive moving charge of $\overrightarrow{dI_2}$ is computed via equation (36).

$$\left|\overrightarrow{dFC_{12+}}\right| = \frac{1}{2}\,\frac{1}{4\,\pi\,\epsilon}\,\frac{dQ_1\,dQ_2}{|\overrightarrow{r}|^2}. \qquad (36)$$

where $dQ_1$ and $dQ_2$are the infinitesimal positive charges that are needed to generate the currents $I_1$ in $\overrightarrow{dI_1}$ and $I_2$ in $\overrightarrow{dI_2}$, respectively. The electric force $\left|\overrightarrow{dFC_{12+}}\right|$ is an attractive force in the direction of the positive $\overrightarrow{u_1}$-axis.

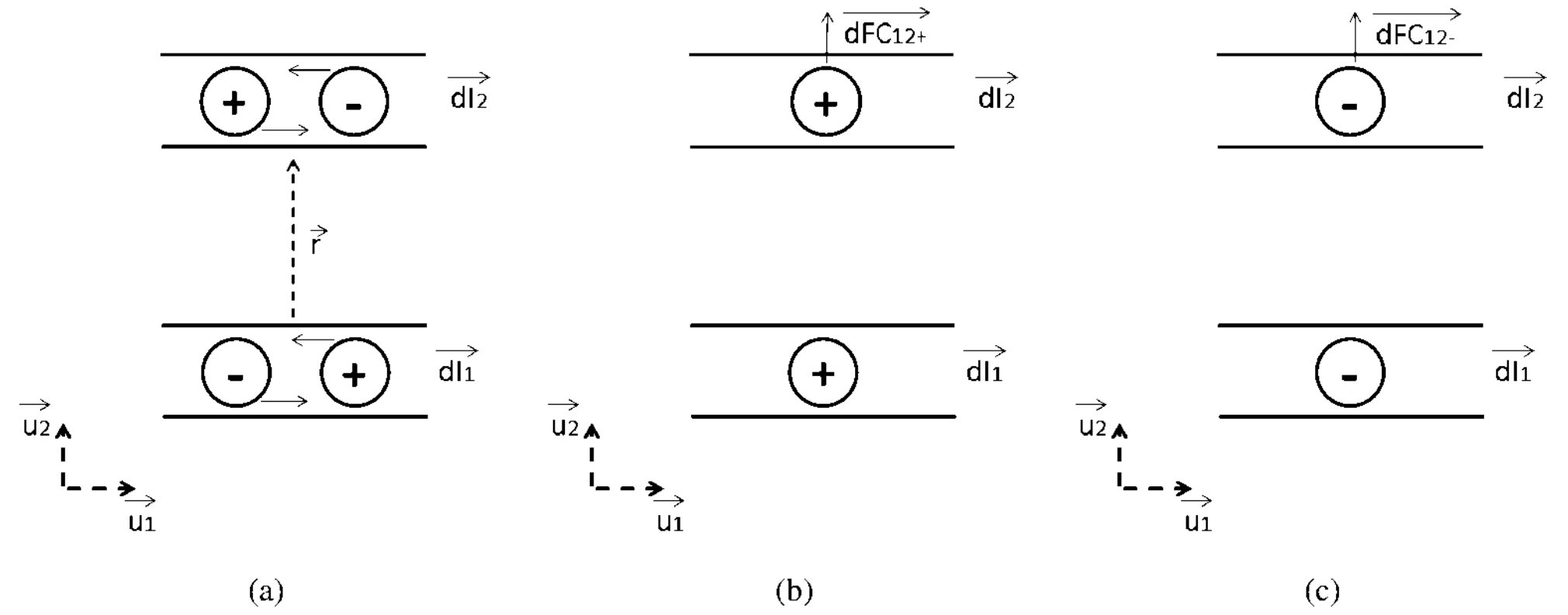

Figure 10: Shows case (1) for $\overrightarrow{dI_1}$ and $\overrightarrow{dI_2}$ with currents that propagate in opposite directions. (a) Two parallel current elements $\overrightarrow{dI_1}$ and $\overrightarrow{dI_2}$ along the $\overrightarrow{u_1}$-axis. (b) The relative positive current charge of $\overrightarrow{dI_1}$ seen by the positive charge of $\overrightarrow{dI_2}$. (c) The relative negative current charge of $\overrightarrow{dI_1}$ seen by the negative charge of $\overrightarrow{dI_2}$.

The magnitude of the electric force $\left|\overrightarrow{dFC_{12-}}\right|$ that the relative positive charge of $\overrightarrow{dI_1}$ affects the negative moving charge of $\overrightarrow{dI_2}$ is computed via equation (37).

$$\left|\overrightarrow{dFC_{12-}}\right| = \frac{1}{2}\,\frac{1}{4\,\pi\,\epsilon}\,\frac{dQ_1\,dQ_2}{|\overrightarrow{r}|^2}. \tag{37}$$

The electric force $\left|\overrightarrow{dFC_{12-}}\right|$ is an attractive force in the direction of the positive $\overrightarrow{u_1}$-axis.

The electric forces $\left|\overrightarrow{dFC_{12+}}\right|$ and $\left|\overrightarrow{dFC_{12-}}\right|$ are along the $\overrightarrow{u_1}$-axis, which is parallel to the current direction where the current charges are moving freely without affecting the current element by any force; refer to section (II-A). Given that, the total electric force that is applied on current element $\overrightarrow{dI_2}$ by its moving charges due to the existence of $\overrightarrow{dI_1}$ is zero, as shown in equation (38).

$$\overrightarrow{dF_{12}} = 0. \tag{38}$$

The electric forces $\left|\overrightarrow{dFC_{12+}}\right|$ and $\left|\overrightarrow{dFC_{12-}}\right|$ do not affect the movement of the charges. This can be explained in part by the fact that these forces have been encountered, i.e., canceled, by the repulsive forces between the current charges and the driving force of the current to maintain the uniform distribution of the charges, to maintain the zero net charge of the current element, to maintain the current value, and to maintain the speed of light for the charges.

For the second situation, by following a similar method as for the previous one, the electric force $\overrightarrow{dF_{12}}$ on $\overrightarrow{dI_2}$ due to $\overrightarrow{dI_1}$ is found when the currents of $\overrightarrow{dI_1}$ and $\overrightarrow{dI_2}$ propagate in opposite directions, as shown in figure (12 a). According to the current charge relativity theory mentioned in section (IV-B), the moving charges of $\overrightarrow{dI_1}$ appear completely positive in the relative frame for the positive moving charges of $\overrightarrow{dI_2}$; refer to figure (12 b). Meanwhile, the moving charges of $\overrightarrow{dI_1}$ appear completely negative in the relative frame for the negative moving charges of $\overrightarrow{dI_2}$; refer to figure (12 c). Thus, the relative charges of $\overrightarrow{dI_1}$ repel the charges of $\overrightarrow{dI_2}$, but these repulsive forces are along the $\overrightarrow{u_1}$-axis, which is parallel to the current direction where the current charges are moving freely without affecting the current element by any force; refer to section (II-A). Given that, the total electric force that is applied on the current element $\overrightarrow{dI_2}$ by its moving charges due to the existence of $\overrightarrow{dI_1}$ is zero, i.e., $\overrightarrow{dF_{12}} = 0$.

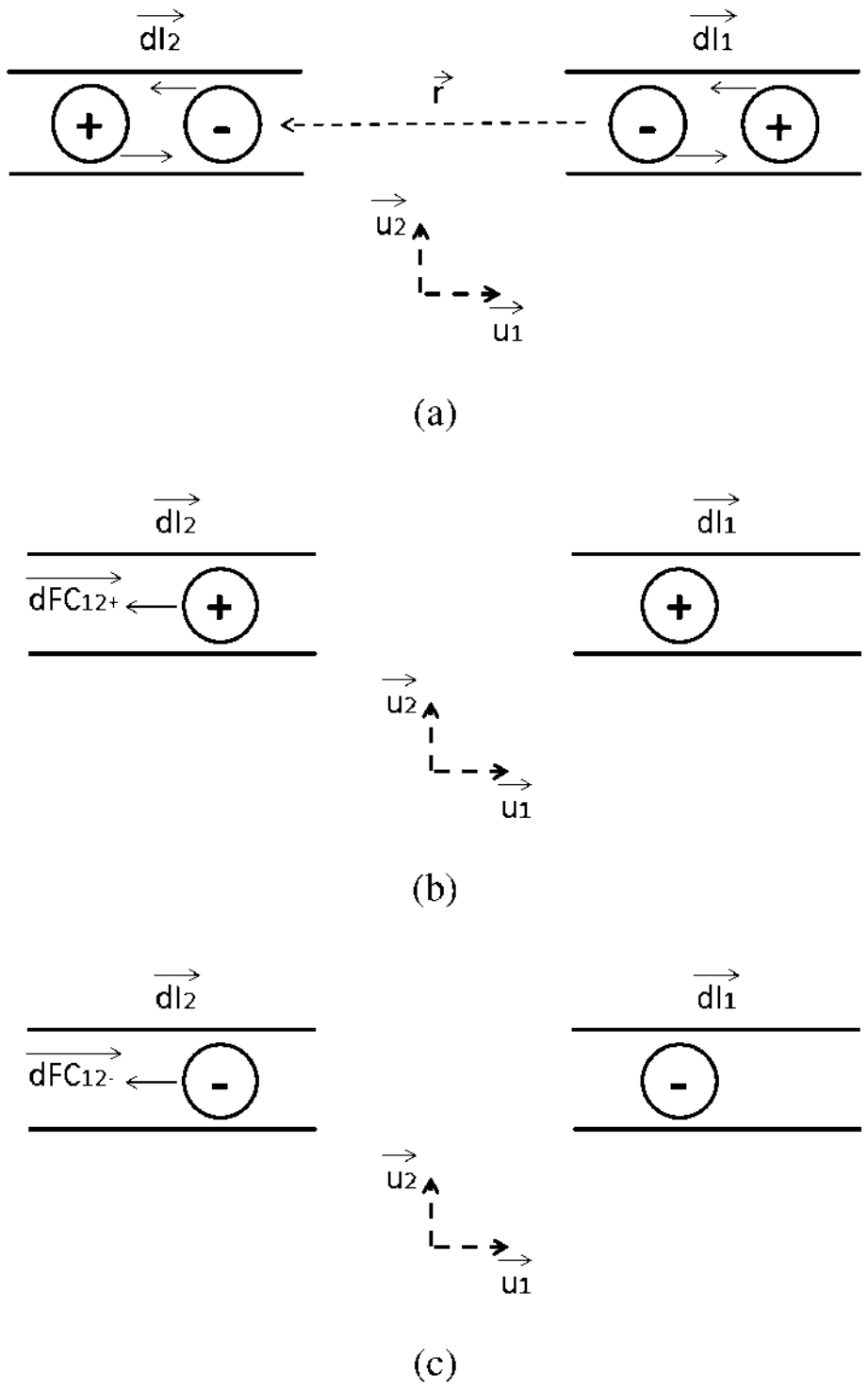

Figure 12: Shows case (2) for $\overrightarrow{dI_1}$ and $\overrightarrow{dI_2}$ with currents that propagate in opposite directions direction. (a) Two parallel current elements $\overrightarrow{dI_1}$ and $\overrightarrow{dI_2}$ with currents that propagate along the $\overrightarrow{u_1}$-axis. (b) The relative positive current charge of $\overrightarrow{dI_1}$ seen by the positive charge of $\overrightarrow{dI_2}$. (c) The relative negative current charge of $\overrightarrow{dI_1}$ seen by the negative charge of $\overrightarrow{dI_2}$.

Using the results for case (1) and case (2), a general expression is written for the electric force law for two parallel current elements at arbitrary positions. The magnitude of the electric force that affects $\overrightarrow{dI_2}$ due to the existence of $\overrightarrow{dI_1}$ is defined in equation (39).

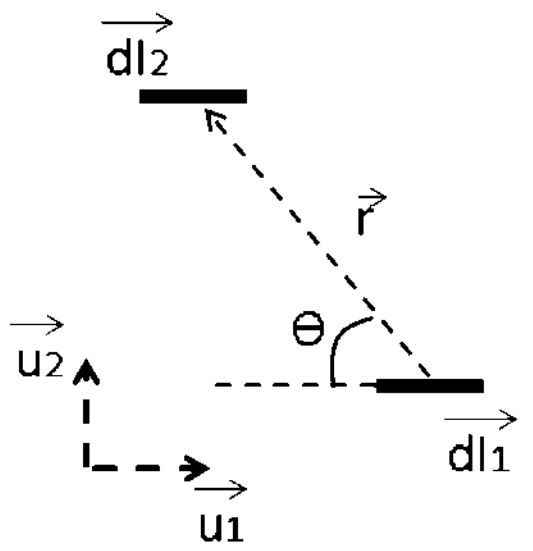

Figure 13: Shows two parallel current elements at arbitrary positions and angle.

$$\left|\overrightarrow{dF_{12}}\right| = \frac{\mu}{4\pi}\frac{I_1 I_2}{|\overrightarrow{r}|^2}\, dl\, dl\, |\overrightarrow{a_1} \times \overrightarrow{a_r}| . \tag{39}$$

where $|\overrightarrow{a_1} \times \overrightarrow{a_r}|$ is the magnitude of the cross-product between $\overrightarrow{a_1}$, i.e., the unit direction of the current propagation for current element $\overrightarrow{dI_1}$, and $\overrightarrow{a_r}$, i.e., the unit direction of the of displacement vector $\overrightarrow{r}$, as shown in figure (13). The magnitude of this cross-product is the sine of the angle $\theta$ between $\overrightarrow{dI_1}$ and $\overrightarrow{r}$, i.e.,$|\overrightarrow{a_1} \times \overrightarrow{a_r}| = \sin\theta$. Equation (39) indicates that only the perpendicular component of forces on the moving charges of $\overrightarrow{dI_2}$ due to the relative charges of $\overrightarrow{dI_1}$ affects the current element $\overrightarrow{dI_2}$, while the horizontal component of these forces, i.e., the component that is along the current propagation in $\overrightarrow{dI_2}$, does not affect $\overrightarrow{dI_2}$.

Equation (39) is updated to include the direction of the electric force on $\overrightarrow{dI_2}$ due to the existence of $\overrightarrow{dI_1}$. The electric force lies on the same plane that contains the current elements $\overrightarrow{dI_1}$ and $\overrightarrow{dI_2}$. The direction of this force is perpendicular to the direction of $\overrightarrow{dI_2}$. This force points toward $\overrightarrow{dI_1}$ if the currents in $\overrightarrow{dI_1}$ and $\overrightarrow{dI_2}$ propagate in the same direction, while this force points away from $\overrightarrow{dI_1}$ if the currents in $\overrightarrow{dI_1}$ and $\overrightarrow{dI_2}$ propagate in opposite directions. The updated equation is shown in equation (40).

$$\overrightarrow{dF_{12}} = \left|\overrightarrow{dF_{12}}\right| \frac{\left(\overrightarrow{a_2} \times \frac{\overrightarrow{a_1}\times\overrightarrow{a_r}}{|\overrightarrow{a_1}\times\overrightarrow{a_r}|}\right)}{\left|\overrightarrow{a_2} \times \frac{\overrightarrow{a_1}\times\overrightarrow{a_r}}{|\overrightarrow{a_1}\times\overrightarrow{a_r}|}\right|} . \tag{40}$$

where $\overrightarrow{a_2}$ is the unit direction of the current element $\overrightarrow{dI_2}$. $\frac{\left(\overrightarrow{a_2} \times \frac{\overrightarrow{a_1}\times\overrightarrow{a_r}}{|\overrightarrow{a_1}\times\overrightarrow{a_r}|}\right)}{\left|\overrightarrow{a_2} \times \frac{\overrightarrow{a_1}\times\overrightarrow{a_r}}{|\overrightarrow{a_1}\times\overrightarrow{a_r}|}\right|}$ expresses the direction of the force. The detailed formula of equation (40) is shown in equation (41).

$$\overrightarrow{dF_{12}} = \frac{\mu}{4\pi}\frac{I_1 I_2}{|\overrightarrow{r}|^2}\, dl\, dl\, |\overrightarrow{a_1} \times \overrightarrow{a_r}| \frac{\left(\overrightarrow{a_2} \times \frac{\overrightarrow{a_1}\times\overrightarrow{a_r}}{|\overrightarrow{a_1}\times\overrightarrow{a_r}|}\right)}{\left|\overrightarrow{a_2} \times \frac{\overrightarrow{a_1}\times\overrightarrow{a_r}}{|\overrightarrow{a_1}\times\overrightarrow{a_r}|}\right|} . \tag{41}$$

Equation (41) is simplified to equation (42).

$$\overrightarrow{dF_{12}} = \frac{\mu}{4\pi}\frac{I_1 I_2}{|\overrightarrow{r}|^2}\, dl\, dl\, \frac{\left(\overrightarrow{a_2} \times \overrightarrow{a_1} \times \overrightarrow{a_r}\right)}{\left|\overrightarrow{a_2} \times \frac{\overrightarrow{a_1}\times\overrightarrow{a_r}}{|\overrightarrow{a_1}\times\overrightarrow{a_r}|}\right|} . \tag{42}$$

Equation (42) is simplified further to equation (43) because $\left|\overrightarrow{a_2} \times \frac{\overrightarrow{a_1}\times\overrightarrow{a_r}}{|\overrightarrow{a_1}\times\overrightarrow{a_r}|}\right| = 1$.

$$\overrightarrow{dF_{12}} = \frac{\mu}{4\pi}\frac{I_1 I_2}{|\overrightarrow{r}|^2}\, dl\, dl\, (\overrightarrow{a_2} \times \overrightarrow{a_1} \times \overrightarrow{a_r}). \tag{43}$$

Equation (43) is rewritten as equation (44).

$$\overrightarrow{dF_{12}} = \frac{\mu}{4\pi}\frac{I_1 I_2}{|\overrightarrow{r}|^2} (dl\, \overrightarrow{a_2} \times (dl\, \overrightarrow{a_1} \times \overrightarrow{a_r})). \tag{44}$$

Equation (44) is rewritten as equation (45).

$$\overrightarrow{dF_{12}} = \frac{\mu}{4\pi}\frac{I_1 I_2}{|\overrightarrow{r}|^2} (\overrightarrow{dl_2} \times (\overrightarrow{dl_1} \times \overrightarrow{a_r})). \tag{45}$$

where $\overrightarrow{dl_1} = dl\, \overrightarrow{a_1}$ is the directional infinitesimal length of the current element $\overrightarrow{dI_1}$, and $\overrightarrow{dl_2} = dl\, \overrightarrow{a_2}$ is the directional infinitesimal length of the current element $\overrightarrow{dI_2}$. Equation (45) is rewritten in terms of the current elements, as in equation (46).

$$\overrightarrow{dF_{12}} = \frac{\mu}{4\pi}\frac{1}{|\overrightarrow{r}|^2} (\overrightarrow{dI_2} \times (\overrightarrow{dI_1} \times \overrightarrow{a_r})). \tag{46}$$

Equation (46) is an exact equivalent, in both magnitude and direction, to the well-known magnetic force law between two parallel filamentary current elements.

*2) Two Perpendicular Filamentary Current Elements on One Plane:* This section derives the electric force law between two perpendicular filamentary current elements that are fully contained by one plane. The force law is derived for case (3) and case (4) of the six cases mentioned earlier, and then the general force law for two perpendicular elements on one plane is found.

For case (3), let $\overrightarrow{dI_1}$ and $\overrightarrow{dI_2}$ be two perpendicular current elements, fully contained in one plane, and $\overrightarrow{dI_1}$ is perpendicular to $\overrightarrow{r}$, i.e., $|\overrightarrow{a_1} \times \overrightarrow{a_r}| = 1$, as shown in figure (14). The electric force on $\overrightarrow{dI_2}$ due to the existence of $\overrightarrow{dI_1}$ is found by analyzing the electric force generated on the moving charges of $\overrightarrow{dI_2}$ due to the movement of the charges of $\overrightarrow{dI_1}$. This force is computed by analyzing the changes in the electric field at $\overrightarrow{dI_2}$ due to the movement of the positive and negative charges of $\overrightarrow{dI_1}$ from one side to another at any moment in the current element and by considering the infinitesimal distance between the charges. In this analysis, the net relative charge is ignored because it is zero. The currents propagate in perpendicular directions; therefore, the charges in $\overrightarrow{dI_2}$ see the positive and negative charges of $\overrightarrow{dI_1}$ cross the space of $\overrightarrow{dI_1}$ at the same speed all the time. This causes an equivalent amount of positive and negative charges to exist in the current elements in all relative frames; thus, the net value of the relative electric charge is zero. Given that, the relative current charge is not considered.

The change in the electric field at $\overrightarrow{dI_2}$ occurs due to the exchange in positions between the positive and negative charges inside the current element $\overrightarrow{dI_1}$ during their movements from one side of a current element to the other side at every moment. To illustrate this change, let $T = \{t_0, t_1, t_2, ...\}$ be a set of consecutive time moments such that $t_{i+1} = t_i + dt$, where $i$ is the index of the moment $t_i$ in the set $T$, and $dt$ is an infinitesimal time. Let $q_+$ be a positive point charge placed in the space at moment $t_0$, as shown in figure (15 a) . The electric field of this charge is pointing outward from that charge in all directions through the space. This electric field indicates the existence of the positive charge $q_+$. If, at moment $t_3$, the positive charge $q_+$ is replaced by a negative charge with equivalent amount $q_-$, then the electric field is inward at the charge location at that moment, as shown in figure (15 d). This change in electric field spreads through the space at the speed of light in all directions and falls off in intensity at $1/r^2$ to indicate the existence of this new placed charge. The falling off in intensity obeys $1/r^2$ because the electric charge in this case is assumed to be either static or moving at a constant velocity, i.e., no acceleration [33], [34]. Then, if this negative charge is replaced again by $q_+$ at moment $t_4$, the electric field is outward at the charge location, and this change in electric field spreads through the space at the speed of light in all directions, as well; see figure (15 e). If this replacement of the charges continues, then the changes in the electric field continue spreading through the space; see figure (15 f). The pattern of these changes in the electric field, i.e., inward-outward or outward-inward, depends on the sequence of the replacement of the charges. In a current element, as shown in figure (16 a) , the positive charges move from right to left, while the negative charges move from left to right to generate a current propagating from right to left. This movement is at the speed of light and is seen by any observer at any position for every moment. For the moment $t_i$, the motion of the charges is modeled as follows. At moment $t_i^-$, the positive charges are in the first half on the right side of the space of the current element moving toward the left half. These charges are modeled as a single positive charge at the center of the right half at $t_i^-$. Meanwhile, the negative charges are in the second half on the left side of the space of the current element, moving toward the right half. These charges are modeled as a single negative charge at the center of the left half at $t_i^-$; see figure (16 a). At moment $t_i^+$, the positive charges move and are in the second half on the left side of the space of the current element, moving to exit the current element from the left. These charges are modeled as a single positive charge at the center of the left half at $t_i^+$. Meanwhile, the negative charges are in the first half on the right side of the space of the current element, moving to exit the current element from the right. These charges are modeled as a single negative charge at the center of the right half at $t_i^+$; see figure (16 b). Therefore, at moment $t_i^-$, an outward electric field is emitted from the positive charge in the right half, and an inward electric field is emitted from the negative charge at the left half; see figure (16 a). Meanwhile, at moment $t_i^+$, an inward electric field is emitted from the negative charge in the right half, and an outward electric field is emitted from the positive charge at the left half; see figure (16 b). These changes spread through the space at the speed of light in all directions and fall off in intensity at $1/r^2$ because the current charges are moving at a constant speed $c$, i.e., there is no acceleration. If this continues for a while, the pattern of the electric field changes is as shown in figure (16 c). This exact pattern must be seen by an observer at any position in the space for every moment; otherwise, a fixed current would have different directions at different locations, and this is not true. For example, let current element $\overrightarrow{dI_2}$ observe a current element $\overrightarrow{dI_1}$, as shown in figure (17 a) . The current of $\overrightarrow{dI_1}$ propagates along the negative $\overrightarrow{u_1}$-axis, while the current of $\overrightarrow{dI_2}$ propagates along the positive $\overrightarrow{u_2}$-axis. At any time $t_i$, the current element $\overrightarrow{dI_2}$ sees and experiences the effect of the positive charge of $\overrightarrow{dI_1}$ moving from right to left and the effect of the negative charge of $\overrightarrow{dI_1}$ moving from left to right. At moment $t_i^-$, the positive charge enters the lower of half of $\overrightarrow{dI_2}$, and the negative charge enters the upper half. At this moment, the element $\overrightarrow{dI_2}$ sees the effect of the electric field generated by the positive charge on the right half of $\overrightarrow{dI_1}$ and by the negative charge on the left side. Thus, the lower half of $\overrightarrow{dI_2}$ has an outward electric field from the positive charge at its right side and an inward electric field from the negative charge on its left side; refer to figure (17 b). Notice that the upper half of $\overrightarrow{dI_2}$ has the electric field pattern of the previous moment $t_{i-1}^+$ because the charges and the change in the electric field move at the speed of light. At moment $t_i^+$, the positive charge enters the upper half of $\overrightarrow{dI_2}$, and the negative charge enters the lower half. At this moment, the element $\overrightarrow{dI_2}$ sees the effect of the electric field generated by the negative charge on the right half of $\overrightarrow{dI_1}$ and by the positive charge on the left side. Thus, the lower half of $\overrightarrow{dI_2}$ has an inward electric field from the negative charge at its right side and an outward electric field from the positive charge on its left side; refer to figure (17 c). Notice that the upper half of $\overrightarrow{dI_2}$ has the electric field pattern at $t_i^-$ because the charges and the change in the electric field move at the speed of light.

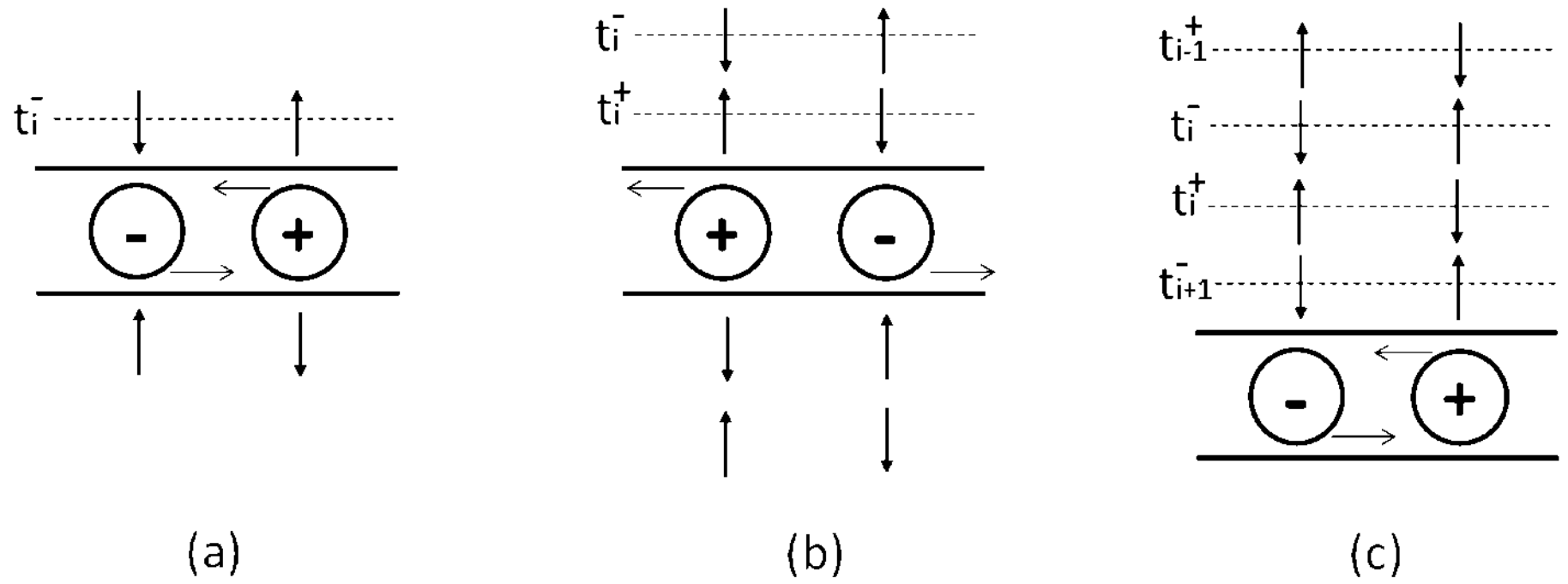

Figure 16: Shows the electric field spreading through the space for a current due to the movement of positive and negative charges. The current propagates from right to left. (a) The location of charges and the electric field at the current element at moment $t_i^-$. (b) The location of charges and the electric field at the current element at moment $t_i^+$. (c) The pattern of the electric field changes spreading through the space due to the movement of charges.

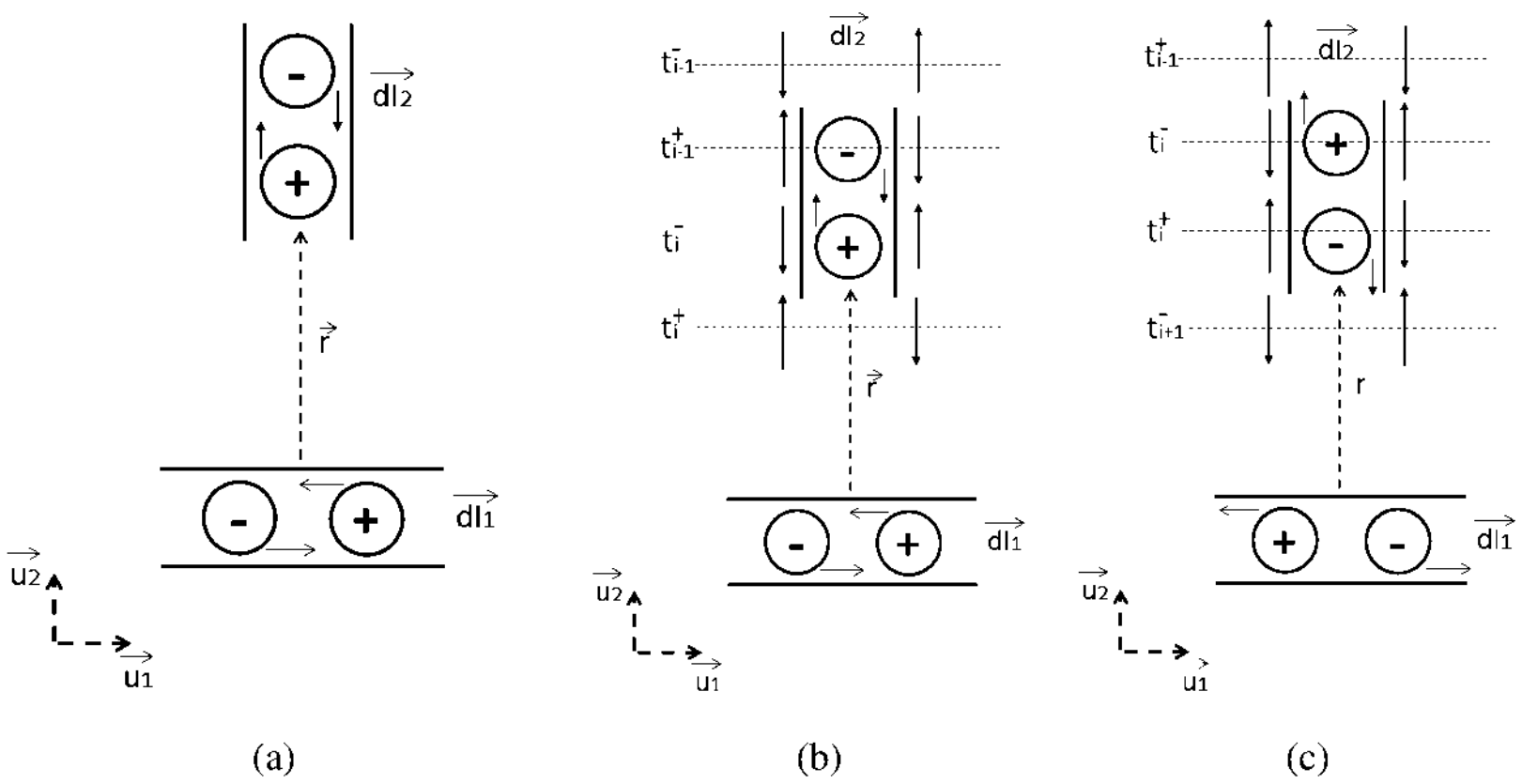

Figure 17: Shows the electric field that affects current element $\overrightarrow{dI_2}$ due to the existence of current element $\overrightarrow{dI_1}$ at time $t_i$. (a) The current elements $\overrightarrow{dI_1}$ and $\overrightarrow{dI_2}$ are perpendicular to each other with currents that propagate toward negative $\overrightarrow{u_1}$ and positive $\overrightarrow{u_2}$, respectively. (b) The electric field seen by $\overrightarrow{dI_2}$ due to the existence of $\overrightarrow{dI_1}$ at moment $t_i^-$. (c) The electric field seen by $\overrightarrow{dI_2}$ due to the existence of $\overrightarrow{dI_1}$ at moment $t_i^+$.

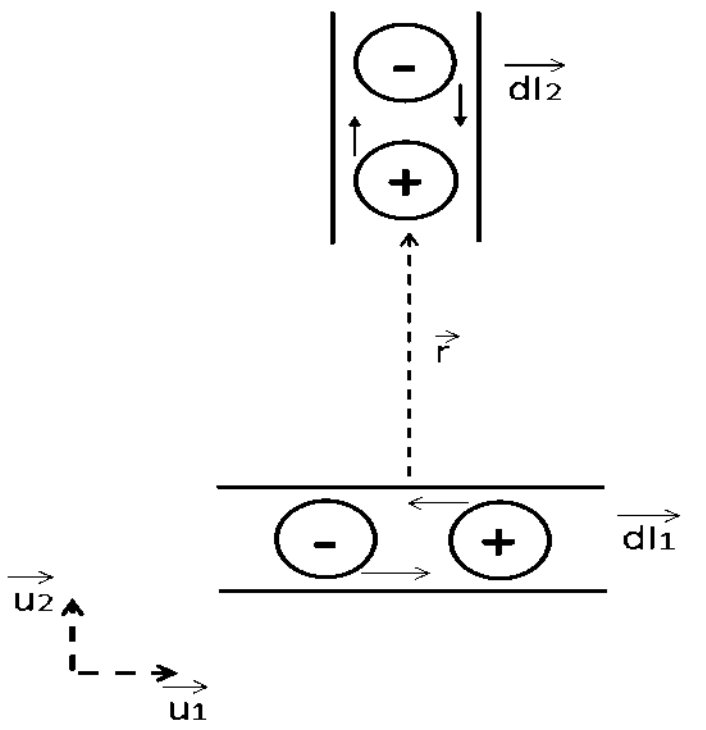

Figure 14: Shows case (3) for two perpendicular current elements that are fully contained in one plane, $\overrightarrow{dI_1}$ and $\overrightarrow{dI_2}$. $\overrightarrow{dI_1}$ is perpendicular to $\overrightarrow{r}$, i.e., $|\overrightarrow{a_1} \times \overrightarrow{a_r}| = 1$. The currents propagate along the negative $\overrightarrow{u_1}$-axis and positive $\overrightarrow{u_2}$-axis for $\overrightarrow{dI_1}$ and $\overrightarrow{dI_2}$, respectively.

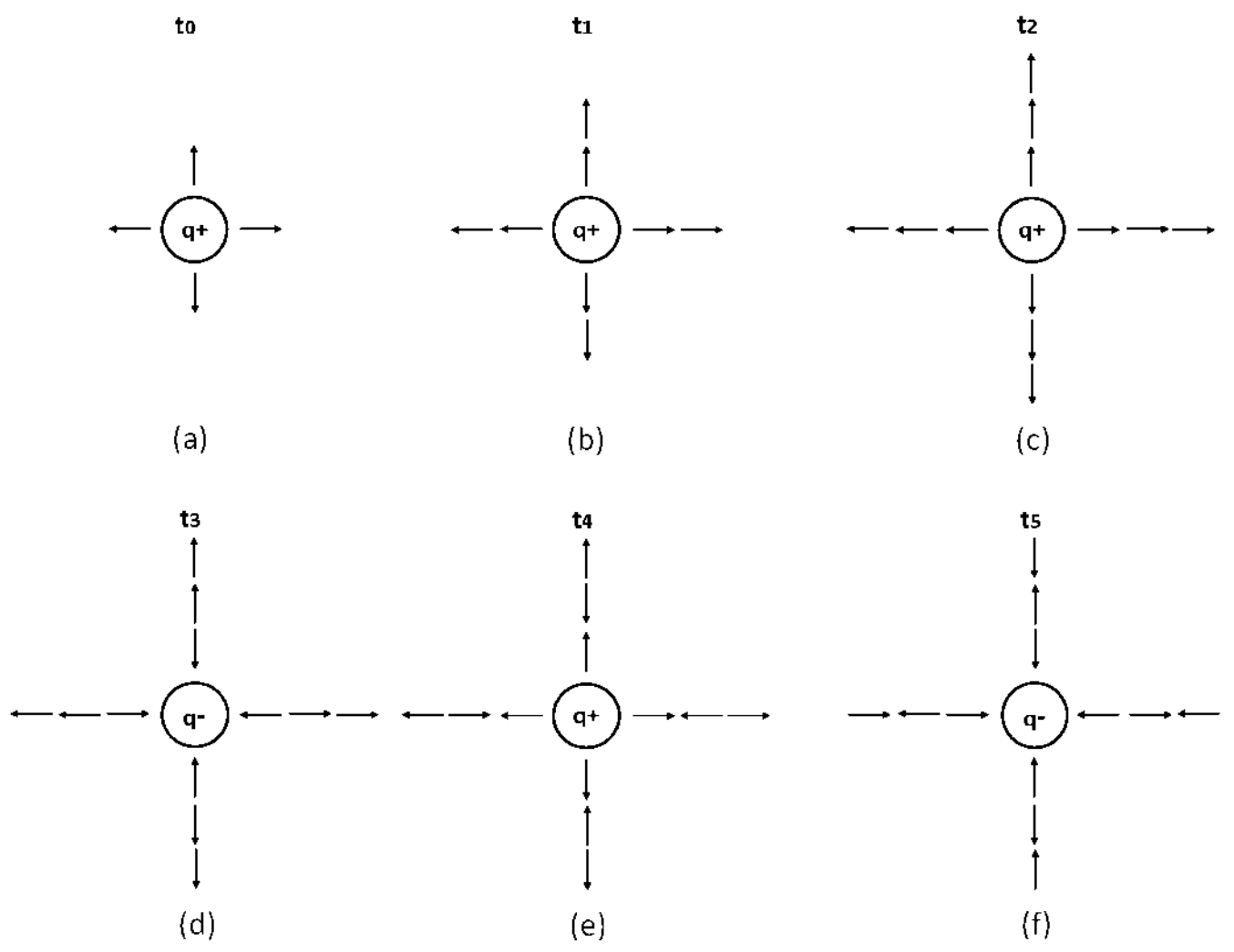

Figure 15: Shows the electric field spreading through the space for a charged object with a charge changing from positive charge $q_+$ to negative charge $q_-$. The positive charge is placed in space at time $t_0$. The charge is changed to negative, positive, and then negative at times $t_3$, $t_4$, and $t_5$, respectively.

The generated pattern of changes indicates discontinuity in the electric field spreading in the space due to the movement of charges in the current element $\overrightarrow{dI_1}$. This discontinuity indicates the existence of an electric charge at a discontinuity location according to Gauss' law assuming constant $\epsilon$ [18]. For example, in figure (17 b), let us examine the four locations of discontinuity of the electric field surrounding the current element $\overrightarrow{dI_2}$, as shown in figure (18 a). At location $A1$, the electric field is along the positive $\overrightarrow{u_2}$-axis, while the electric field beneath it is along the negative $\overrightarrow{u_2}$-axis; this then indicates the existence of a positive charge at $A1$ [35], i.e., $\nabla \cdot E(A1) > 0$, where $\nabla\cdot$ is the divergence operation and $E(A1)$ is the electric field at $A1$. This charge has an electric field of $E(A1)$ surrounding it. At location $A2$, the electric field is along the negative $\overrightarrow{u_2}$-axis, while the electric field beneath it is along the positive $\overrightarrow{u_2}$-axis; this then indicates the existence of a negative charge at $A2$, i.e., $\nabla \cdot E(A2) < 0$. This charge has an electric field of $E(A2)$ surrounding it. At location $A3$, the electric field is along the negative $\overrightarrow{u_2}$-axis, while the electric field beneath it is along the positive $\overrightarrow{u_2}$-axis; this then indicates the existence of a negative charge at $A3$, i.e., $\nabla \cdot E(A3) < 0$. This charge has an electric field of $E(A3)$ surrounding it. At location $A4$, the electric field is along the positive $\overrightarrow{u_2}$-axis, while the electric field beneath it is along the negative $\overrightarrow{u_2}$-axis; this then indicates the existence of a positive charge at $A4$, i.e., $\nabla \cdot E(A4) > 0$. This charge has an electric field of $E(A4)$ surrounding it. These charges are shown in figure (18 b). Following a similar method, these charges are found for the electric field pattern at $t_i^+$ as shown in figures (17 c and 18 c). There is a negative charge at location $A1$, a positive charge at location $A2$, a positive charge at location $A3$, and a negative charge at location $A4$; refer to figure (18 d). These surrounding charges are referred to as discontinuity charges. These charges can be explained in part as a polarization effect of the medium that generates infinitesimal point dipoles opposing each other with finite charges [1], [36]. In a vacuum, the discontinuity charge effect can be explained by the photons that are traveling through the space at the speed of light to indicate changes in the electric field [32], [23]. These photons are able to exert electrical forces on charged particles [37], [38]. Moreover, Altschul in [39] assumed the existence of photons with positive and negative charges. He analyzed radio waves from distant galaxies to obtain a new upper bound on the electrical charge of the photon. He found that if different photons can carry different charges then their upper bound will be at $10^{-46}e$ level, where $e$ is the charge of an electron. Further investigation is needed to explore the nature of the electric force, the nature of the electric charge, and the effect of charge movements into space to explain the existence of the discontinuity charge effect, because there is no apparent source for it in a vacuum despite the indication of its existence by Gauss' law. This investigation may need to be conducted in connection with other explorations to the nature of gravity and nuclear forces within a unified framework, e.g., string theory [40], [41]. Such an investigation is not part of this work and is better suited for future research.

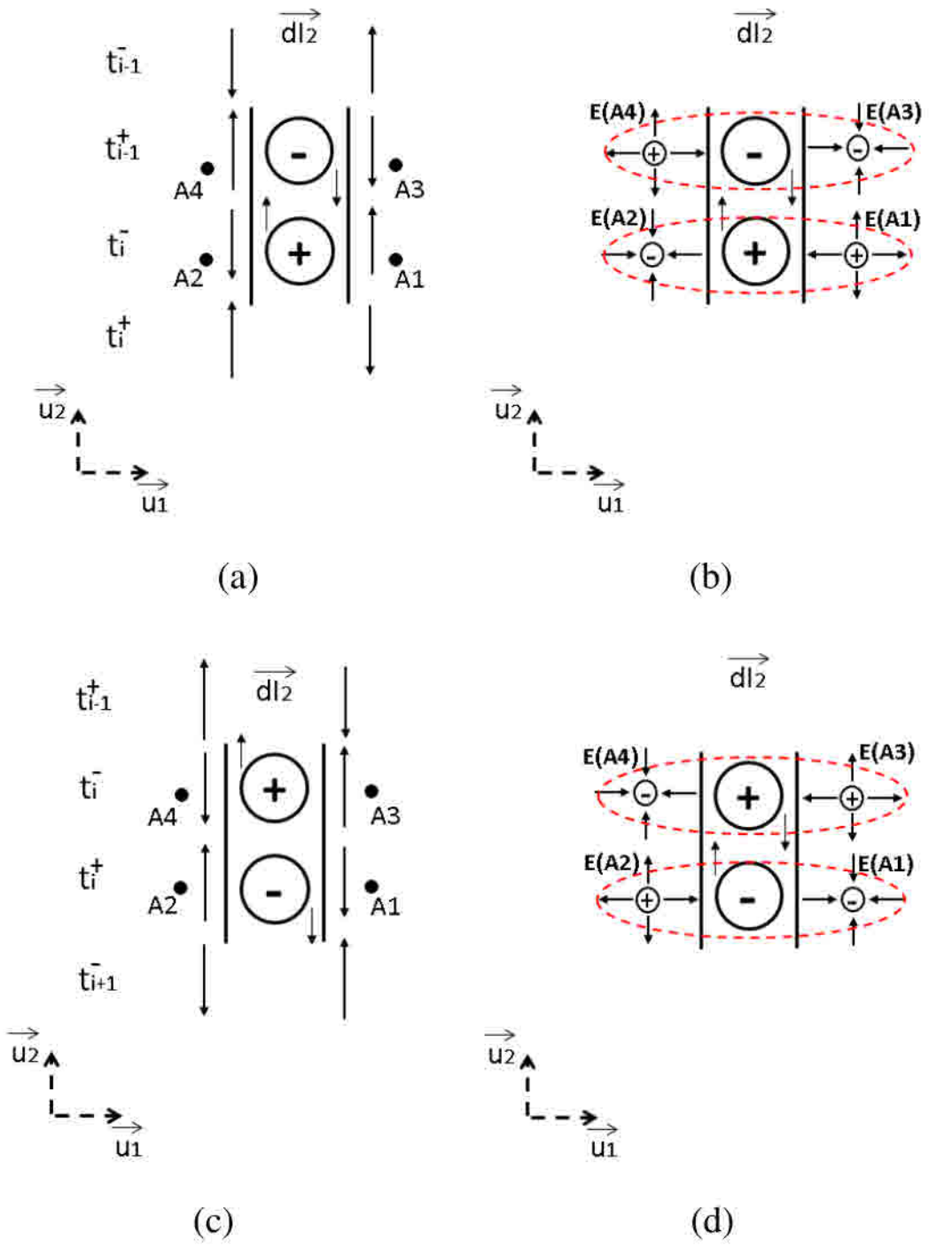

Figure 18: Shows the four discontinuity locations, i.e., $A1$, $A2$, $A3$, and $A4$, and the discontinuity charges surrounding the current element $\overrightarrow{dI_2}$ at time $t_i$. (a) The electric field surrounding $\overrightarrow{dI_2}$ at time $t_i^-$. (b) The discontinuity charges with their electric fields surrounding $\overrightarrow{dI_2}$ at time $t_i^-$. (c) The electric field surrounding $\overrightarrow{dI_2}$ at time $t_i^+$. (d) The discontinuity charges with their electric fields surrounding $\overrightarrow{dI_2}$ at time $t_i^+$. $E(Ai)$ is the electric field at location $Ai$.

The discontinuity charges surrounding the current element $\overrightarrow{dI_2}$ generate an electric force on the charges of that current element. Each of these discontinuity charges emanates an electric field around it that has a magnitude computed by equation (47).

$$\left|\overrightarrow{dE_{12}}(r)\right| = \frac{1}{4\,\pi\,\epsilon}\frac{\left|\frac{dQ_1}{2}\right|}{|\overrightarrow{r}|^2}. \tag{47}$$

where $\left|\overrightarrow{dE_{12}}(r)\right|$ is the magnitude of the electric field from the positive or negative charges of $\overrightarrow{dI_1}$ at $\overrightarrow{dI_2}$. Let the direction of the current in $\overrightarrow{dI_2}$ be as shown in figure (18 ). Then, at $t_i^-$, the positive discontinuity charge at location $A1$ produces a repulsive force on the positive charge of $\overrightarrow{dI_2}$ toward the negative $\overrightarrow{u_1}$-axis, as defined in equation (48).

$$\overrightarrow{dFC^{i-}_{A12+}} = -\left|\overrightarrow{dE_{12}}(r)\right|\frac{dQ_2}{2}\,\overrightarrow{u_1}. \tag{48}$$

where $\overrightarrow{dFC^{i-}_{A12+}}$ is the electric force on the positive charge of $\overrightarrow{dI_2}$ due to the discontinuity charge at $A1$ at moment $t_i^-$. In a similar manner, the negative discontinuity charge at location $A2$ produces an attractive force on the positive charge of $\overrightarrow{dI_2}$ toward the negative $\overrightarrow{u_1}$-axis, as defined in equation (49).

$$\overrightarrow{dFC^{i-}_{A22+}} = -\left|\overrightarrow{dE_{12}}(r)\right|\frac{dQ_2}{2}\,\overrightarrow{u_1}. \tag{49}$$

where $\overrightarrow{dFC^{i-}_{A22+}}$ is the electric force on the positive charge of $\overrightarrow{dI_2}$ due to the discontinuity charge at $A2$ at moment $t_i^-$. The negative discontinuity charge at location $A3$ produces a repulsive force on the negative charge of $\overrightarrow{dI_2}$ toward the negative $\overrightarrow{u_1}$-axis, as defined in equation (50).

$$\overrightarrow{dFC^{i-}_{A32-}} = -\left|\overrightarrow{dE_{12}}(r)\right|\frac{dQ_2}{2}\,\overrightarrow{u_1}. \tag{50}$$

where $\overrightarrow{dFC^{i-}_{A32-}}$ is the electric force on the negative charge of $\overrightarrow{dI_2}$ due to the discontinuity charge at $A3$ at moment $t_i^-$. The positive discontinuity charge at location $A4$ produces an attractive force on the negative charge of $\overrightarrow{dI_2}$ toward the negative $\overrightarrow{u_1}$-axis, as defined in equation (51).

$$\overrightarrow{dFC^{i-}_{A42-}} = -\left|\overrightarrow{dE_{12}}(r)\right|\frac{dQ_2}{2}\,\overrightarrow{u_1}. \tag{51}$$

where $\overrightarrow{dFC^{i-}_{A42-}}$ is the electric force on the negative charge of $\overrightarrow{dI_2}$ due to the discontinuity charge at $A4$ at moment $t_i^-$.

The total electric force on the positive charge of $\overrightarrow{dI_2}$ is the sum of $\overrightarrow{dFC^{i-}_{A12+}}$ and $\overrightarrow{dFC^{i-}_{A22+}}$, and it is computed by equation (52).

$$\overrightarrow{dFC^{i-}_{12+}} = -\left|\overrightarrow{dE_{12}}(r)\right| \, dQ_2 \, \overrightarrow{u_1}. \quad (52)$$

where $\overrightarrow{dFC^{i-}_{12+}}$ is the total electric force on the positive charge of $\overrightarrow{dI_2}$ due to the existence of $\overrightarrow{dI_1}$ at $t^-_i$.

The total electric force on the negative charge of $\overrightarrow{dI_2}$ is the sum of $\overrightarrow{dFC^{i-}_{A32-}}$ and $\overrightarrow{dFC^{i-}_{A42-}}$, and it is computed by equation (53).

$$\overrightarrow{dFC^{i-}_{12-}} = -\left|\overrightarrow{dE_{12}}(r)\right| \, dQ_2 \, \overrightarrow{u_1}. \quad (53)$$

where $\overrightarrow{dFC^{i-}_{12-}}$ is the total electric force on the negative charge of $\overrightarrow{dI_2}$ due to the existence of $\overrightarrow{dI_1}$ at $t^-_i$.

The electric forces $\overrightarrow{dFC^{i-}_{12+}}$ and $\overrightarrow{dFC^{i-}_{12-}}$ on the crossing charges of $\overrightarrow{dI_2}$ are perpendicular to their motion, and these charges are not permitted to leave their filamentary current element. Given that, these charges push the filamentary current element by these forces; refer to section (II-A).

The magnitude of the total electric force that is applied on the current element $\overrightarrow{dI_2}$ due to the existence of the current element $\overrightarrow{dI_1}$ at $t^-_i$ is computed as shown in equation (54).

$$\overrightarrow{dF^{i-}_{12}} = \overrightarrow{dFC^{i-}_{12+}} + \overrightarrow{dFC^{i-}_{12-}},$$

$$\overrightarrow{dF^{i-}_{12}} = -2\left|\overrightarrow{dE_{12}}(r)\right| \, dQ_2 \, \overrightarrow{u_1}. \quad (54)$$

where $\overrightarrow{dF^{i-}_{12}}$ is the force applied on $\overrightarrow{dI_2}$ due to the existence of $\overrightarrow{dI_1}$ at $t^-_i$. By substituting equation (47) for $\left|\overrightarrow{dE_{12}}(r)\right|$, equation (54) is rewritten as equation (55).

$$\overrightarrow{dF^{i-}_{12}} = -\frac{1}{4\,\pi\,\epsilon}\frac{dQ_1\,dQ_2}{|\overrightarrow{r}|^2}\,\overrightarrow{u_1}. \quad (55)$$

Following a similar method, the force that is applied on $\overrightarrow{dI_2}$ due to the existence of $\overrightarrow{dI_1}$ at $t^+_i$, as shown in figure (18 d), is computed via equation (56).

$$\overrightarrow{dF^{i+}_{12}} = -\frac{1}{4\,\pi\,\epsilon}\frac{dQ_1\,dQ_2}{|\overrightarrow{r}|^2}\,\overrightarrow{u_1}. \quad (56)$$

where $\overrightarrow{dF^{i+}_{12}}$ is the force applied on $\overrightarrow{dI_2}$ due to the existence of $\overrightarrow{dI_1}$ at $t^+_i$. The force $\overrightarrow{dF^{i+}_{12}}$ is equivalent in both magnitude and direction to the force $\overrightarrow{dF^{i-}_{12}}$. Then, the force that is applied on $\overrightarrow{dI_2}$ at moment $t_i$ for $dt$ is computed via equation (57).

$$\overrightarrow{dF_{12}}\,dt = \overrightarrow{dF^{i-}_{12}}\,\frac{dt}{2} + \overrightarrow{dF^{i+}_{12}}\,\frac{dt}{2},$$

$$\overrightarrow{dF_{12}} = -\frac{1}{4\,\pi\,\epsilon}\frac{dQ_1\,dQ_2}{|\overrightarrow{r}|^2}\,\overrightarrow{u_1}. \quad (57)$$

where $\overrightarrow{dF_{12}}$ is the total electric force that is applied on $\overrightarrow{dI_2}$ by the perpendicular forces applied on its moving charges due to the existence of $\overrightarrow{dI_1}$. Notice that the time index is removed because this force is applied on $\overrightarrow{dI_2}$ all of the time as long as the current elements $\overrightarrow{dI_1}$ and $\overrightarrow{dI_2}$ do not change. This force is in the direction of the negative $\overrightarrow{u_1}$-axis.

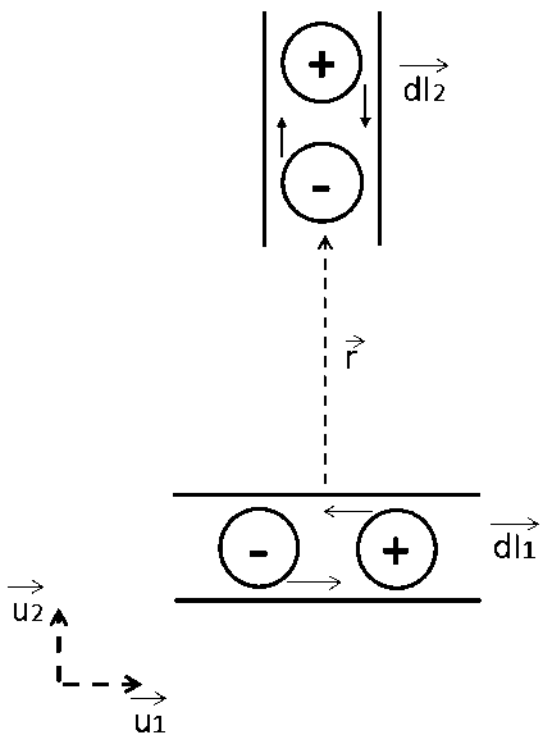

Figure 19: Shows case (3) for two perpendicular current elements that are fully contained in one plane, $\overrightarrow{dI_1}$ and $\overrightarrow{dI_2}$. $\overrightarrow{dI_1}$ is perpendicular to $\overrightarrow{r}$, i.e., $|\overrightarrow{a_1} \times \overrightarrow{a_r}| = 1$. The currents propagate along the negative $\overrightarrow{u_1}$-axis and negative $\overrightarrow{u_2}$-axis for $\overrightarrow{dI_1}$ and $\overrightarrow{dI_2}$, respectively.

The total electric force on $\overrightarrow{dI_2}$ due to $\overrightarrow{dI_1}$ is rewritten in terms of the current flowing through the current elements by multiplying and dividing by $c^2$ on the right side of equation (57), as shown in equation (58).

$$\overrightarrow{dF_{12}} = -\frac{c^2}{c^2}\frac{1}{4\,\pi\,\epsilon}\frac{dQ_1\,dQ_2}{|\overrightarrow{r}|^2}\,\overrightarrow{u_1}. \quad (58)$$

Equation (58) is rewritten and simplified further as shown in equations (59 and 60).

$$\overrightarrow{dF_{12}} = -\frac{1}{4\,\pi\,(\epsilon\,c^2)}\frac{(Q_1\,ds\,c)\,(Q_2\,ds\,c)}{|\overrightarrow{r}|^2}\,dl\,dl\,\overrightarrow{u_1}. \quad (59)$$

$$\overrightarrow{dF_{12}} = -\frac{1}{4\,\pi\,(\epsilon\,c^2)}\frac{I_1\,I_2}{|\overrightarrow{r}|^2}\,dl\,dl\,\overrightarrow{u_1}. \quad (60)$$

Equation (60) is simplified further as shown in equation (61) because $\mu = \frac{1}{\epsilon\,c^2}$, where $\mu$ is the magnetic permeability.

$$\overrightarrow{dF_{12}} = -\frac{\mu}{4\,\pi}\frac{I_1\,I_2}{|\overrightarrow{r}|^2}\,dl\,dl\,\overrightarrow{u_1}. \quad (61)$$

The electric force $\overrightarrow{dF_{12}}$ on $\overrightarrow{dI_2}$ due to $\overrightarrow{dI_1}$, as shown in equation (61), is equivalent in magnitude and direction to the magnetic force between two perpendicular infinitesimal current elements having currents that propagate in the directions shown in figure (14).

For the situation where the current of $\overrightarrow{dI_2}$ is propagating in the opposite direction, as shown in figure (19) , the electric force is computed by equation (62).

$$\overrightarrow{dF_{12}} = \frac{\mu}{4\,\pi}\frac{I_1\,I_2}{|\overrightarrow{r}|^2}\,dl\,dl\,\overrightarrow{u_1}. \quad (62)$$

In this situation, the force is equivalent in magnitude to the force computed for the previous situation described in figure (18), but this force is in the opposite direction. This behavior is similar to the magnetic force law behavior when one of the currents changes its direction to the opposite one.

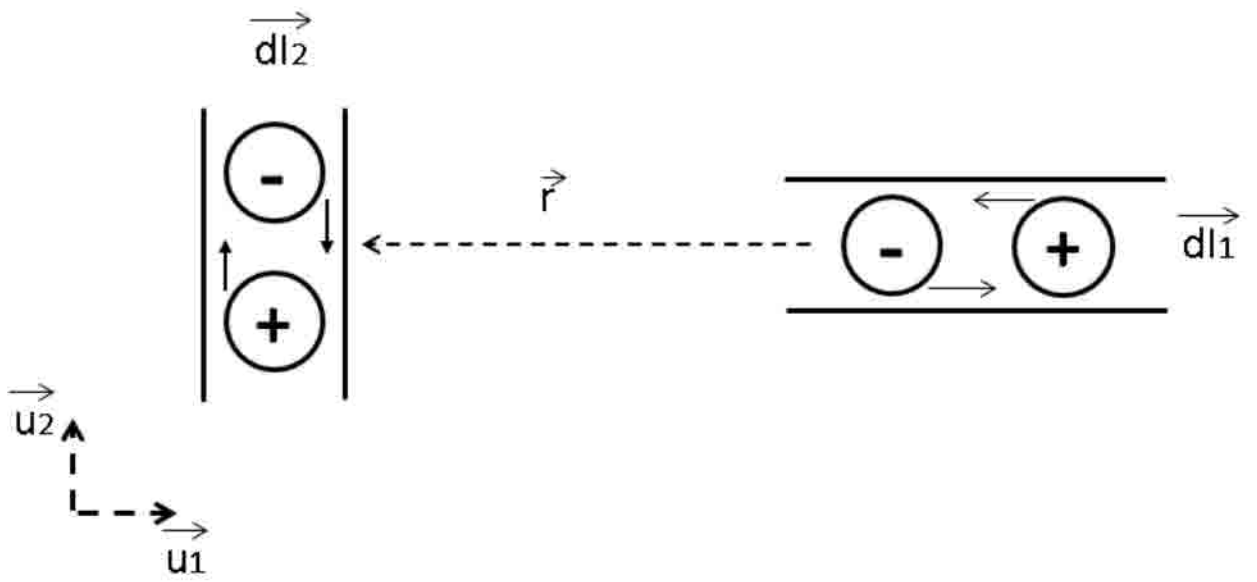

Figure 20: Shows case (4) for two perpendicular current elements that are fully contained in one plane, $\overrightarrow{dI_1}$ and $\overrightarrow{dI_2}$. $\overrightarrow{dI_1}$ is parallel to $\overrightarrow{r}$, i.e., $\overrightarrow{a_1} \times \overrightarrow{a_r} = 0$. The currents propagate along the negative $\overrightarrow{u_1}$-axis and positive $\overrightarrow{u_2}$-axis for $\overrightarrow{dI_1}$ and $\overrightarrow{dI_2}$, respectively.

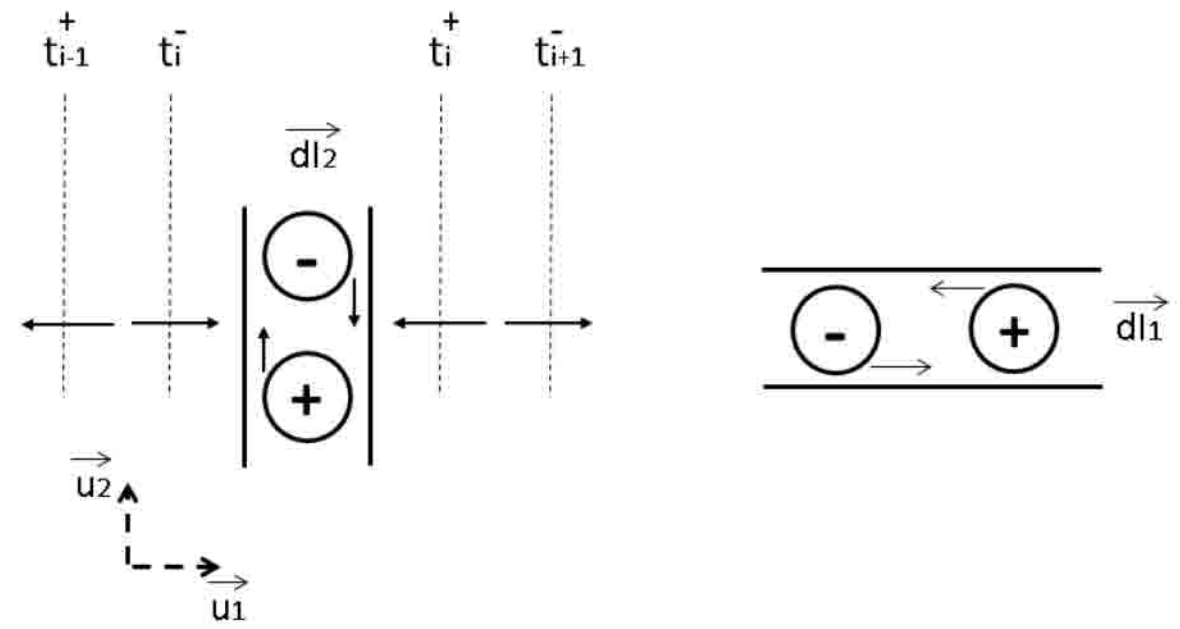

(a)

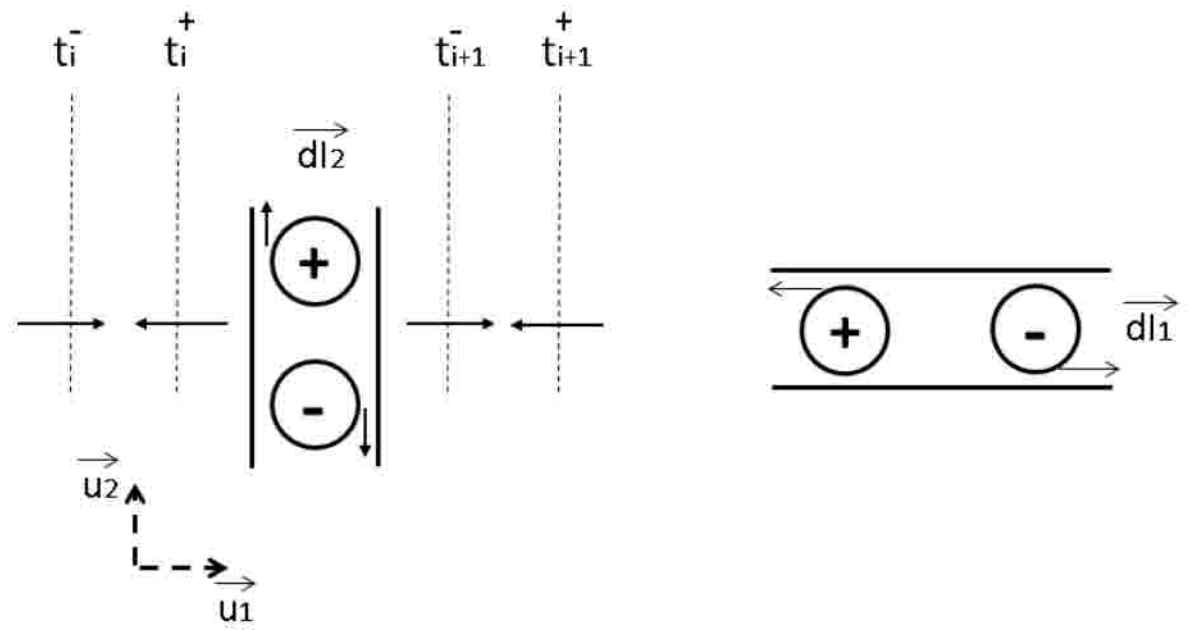

(b)

Figure 21: Shows the electric field surrounding the current element $\overrightarrow{dI_2}$ due to the existence of current element $\overrightarrow{dI_1}$ at time $t_i$. (a) The electric field seen by $\overrightarrow{dI_2}$ due to the existence of $\overrightarrow{dI_1}$ at moment $t_i^-$. (b) The electric field seen by $\overrightarrow{dI_2}$ due to the existence of $\overrightarrow{dI_1}$ at moment $t_i^+$.

For case (4), $\overrightarrow{dI_2}$ and $\overrightarrow{dI_1}$ are perpendicular to each other and fully contained in one plane, and $\overrightarrow{dI_1}$ is parallel to $\overrightarrow{r}$, i.e., $\overrightarrow{a_1} \times \overrightarrow{a_r} = 0$; see figure (20). Following the same analysis described for case (3), at moment $t_i^-$, the element $\overrightarrow{dI_2}$ sees a positive charge on the right half of $\overrightarrow{dI_1}$ and a negative charge on the left side. Thus, $\overrightarrow{dI_2}$ has an inward electric field from the negative charge on its left side and an outward electric field from the positive charge on its right side from moment $t_i^+$; refer to figure (21 a). At moment $t_i^+$, the element $\overrightarrow{dI_2}$ sees a negative charge on the right half of $\overrightarrow{dI_1}$ and a positive charge on the left side. Thus, $\overrightarrow{dI_2}$ has an outward electric field from the positive charge on its left side and an inward electric field from the negative charge on its right side from moment $t_{i+1}^-$; refer to figure (21 b).

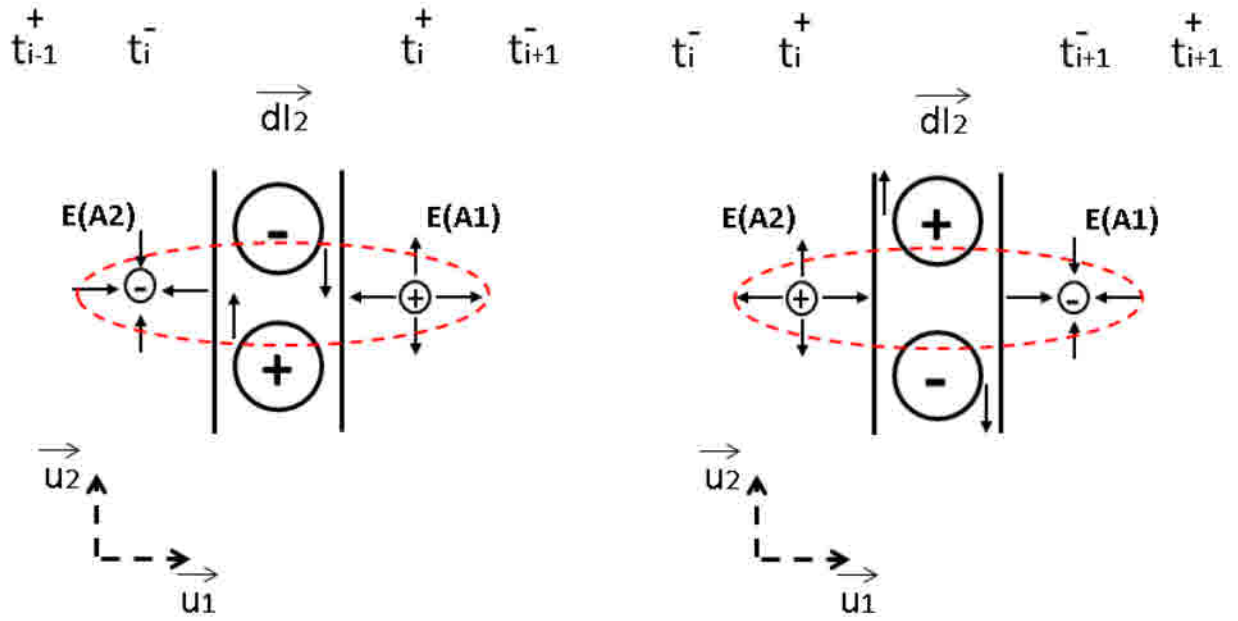

(a) (b)

Figure 22: Shows the discontinuity charges surrounding the current element $\overrightarrow{dI_2}$ due to the existence of current element $\overrightarrow{dI_1}$ at time $t_i$. (a) The discontinuity charges with their electric fields surrounding $\overrightarrow{dI_2}$ at time $t_i^-$. (b) The discontinuity charges with their electric fields surrounding $\overrightarrow{dI_2}$ at time $t_i^+$. $E(A_i)$ is the electric field at location $A_i$.

The generated pattern of changes indicates that discontinuity in the electric field is spreading in the space due to the movement of charges in the current element $\overrightarrow{dI_1}$. This discontinuity indicates the existence of an electric charge at a discontinuity location according to Gauss' law. The generated pattern of the electric field at $t_i^-$ indicates a negative discontinuity charge to the left side of $\overrightarrow{dI_2}$ and a positive discontinuity charge on the right side of $\overrightarrow{dI_2}$; see figure (22 a). Meanwhile, the generated pattern of the electric field at $t_i^+$ indicates a positive discontinuity charge on the left side of $\overrightarrow{dI_2}$ and a negative discontinuity on the right side of $\overrightarrow{dI_2}$; see figure (22 b). The discontinuity charges surrounding the current element $\overrightarrow{dI_2}$ generate an electric force on the charges of that current element. Each of these discontinuity charges emanates an electric field around it that has a magnitude computed by equation (47).

The force produced on $\overrightarrow{dI_2}$ at $t_i^-$ is computed as follows. The negative discontinuity charge on the left side of $\overrightarrow{dI_2}$ produces a repulsive force on the negative charge under effect of $\overrightarrow{dI_2}$ toward the positive $\overrightarrow{u_1}$-axis and produces an attractive force of the same amount on the positive charge under effect of $\overrightarrow{dI_2}$ toward the negative $\overrightarrow{u_1}$-axis. Meanwhile, the positive discontinuity charge on the right side of $\overrightarrow{dI_2}$ produces an attractive force on the negative charge under effect of $\overrightarrow{dI_2}$ toward the positive $\overrightarrow{u_1}$-axis and produces a repulsive force of the same amount on the positive charge under effect of $\overrightarrow{dI_2}$ toward the negative $\overrightarrow{u_1}$-axis. These forces are in opposite directions and are perpendicular to the direction of motion for the charges of $\overrightarrow{dI_2}$. These charges are not permitted to leave their filamentary current element. Given that, these charges push the filamentary current element by these forces. However, these forces have the same magnitude but in opposite directions; therefore, the net force applied on $\overrightarrow{dI_2}$ at $t_i^-$ is zero;

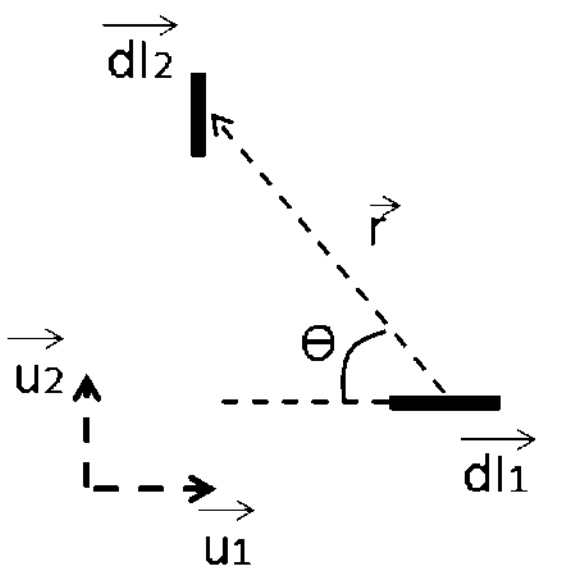

Figure 23: Shows two perpendicular current elements fully contained in one plane at arbitrary positions and angle.

refer to equation (63).

$$\overrightarrow{dF_{12}^{-}} = 0. \tag{63}$$

Following a similar method, the net force produced on $\overrightarrow{dI_2}$ at $t_i^{+}$ is zero; refer to equation (64).

$$\overrightarrow{dF_{12}^{+}} = 0. \tag{64}$$

Given that, the total electric force that is applied on the current element $\overrightarrow{dI_2}$ by its moving charges due to the existence of $\overrightarrow{dI_1}$ is zero, as shown in equation (65).

$$\overrightarrow{dF_{12}} = 0. \tag{65}$$

Using the results for case (3) and case (4), a general expression is written for the electric force law for two perpendicular current elements that lay on one plane at arbitrary positions. The magnitude of the electric force that affects $\overrightarrow{dI_2}$ due to the existence of $\overrightarrow{dI_1}$ is defined in equation (66).

$$\left|\overrightarrow{dF_{12}}\right| = \frac{\mu}{4\,\pi}\frac{I_1\,I_2}{|\overrightarrow{r}|^2}\,dl\,dl\,\left|\overrightarrow{a_1}\times\overrightarrow{a_r}\right|. \tag{66}$$

where $|\overrightarrow{a_1}\times\overrightarrow{a_r}|$ is the magnitude of the cross-product between $\overrightarrow{a_1}$, i.e., the unit direction of the current propagation for current element $\overrightarrow{dI_1}$, and $\overrightarrow{a_r}$, i.e., the unit direction of the of displacement vector $\overrightarrow{r}$, as shown in figure (23). The magnitude of this cross-product is the sine of the angle $\theta$ between $\overrightarrow{dI_1}$ and $\overrightarrow{r}$, i.e.,$|\overrightarrow{a_1}\times\overrightarrow{a_r}| = \sin\theta$.

Equation (66) is updated to include the direction of the electric force on $\overrightarrow{dI_2}$ due to the existence of $\overrightarrow{dI_1}$. The electric force lies on the same plane that contains the current elements $\overrightarrow{dI_1}$ and $\overrightarrow{dI_2}$. The direction of this force is perpendicular to the direction of $\overrightarrow{dI_2}$. The updated equation is shown in equation (67).

$$\overrightarrow{dF_{12}} = \left|\overrightarrow{dF_{12}}\right| \frac{(\overrightarrow{a_2}\times\frac{\overrightarrow{a_1}\times\overrightarrow{a_r}}{|\overrightarrow{a_1}\times\overrightarrow{a_r}|})}{\left|\overrightarrow{a_2}\times\frac{\overrightarrow{a_1}\times\overrightarrow{a_r}}{|\overrightarrow{a_1}\times\overrightarrow{a_r}|}\right|}. \tag{67}$$

where $\overrightarrow{a_2}$ is the unit direction of the current element $\overrightarrow{dI_2}$. $\frac{(\overrightarrow{a_2}\times\frac{\overrightarrow{a_1}\times\overrightarrow{a_r}}{|\overrightarrow{a_1}\times\overrightarrow{a_r}|})}{\left|\overrightarrow{a_2}\times\frac{\overrightarrow{a_1}\times\overrightarrow{a_r}}{|\overrightarrow{a_1}\times\overrightarrow{a_r}|}\right|}$ expresses the direction of the force. The detailed formula of equation (67) is shown in equation (68).

$$\overrightarrow{dF_{12}} = \frac{\mu}{4\,\pi}\frac{I_1\,I_2}{|\overrightarrow{r}|^2}\,dl\,dl\,\left|\overrightarrow{a_1}\times\overrightarrow{a_r}\right| \frac{(\overrightarrow{a_2}\times\frac{\overrightarrow{a_1}\times\overrightarrow{a_r}}{|\overrightarrow{a_1}\times\overrightarrow{a_r}|})}{\left|\overrightarrow{a_2}\times\frac{\overrightarrow{a_1}\times\overrightarrow{a_r}}{|\overrightarrow{a_1}\times\overrightarrow{a_r}|}\right|}. \tag{68}$$

Equation (68) is simplified to equation (69).

$$\overrightarrow{dF_{12}} = \frac{\mu}{4\,\pi}\frac{I_1\,I_2}{|\overrightarrow{r}|^2}\,dl\,dl\,\frac{(\overrightarrow{a_2}\times\overrightarrow{a_1}\times\overrightarrow{a_r})}{\left|\overrightarrow{a_2}\times\frac{\overrightarrow{a_1}\times\overrightarrow{a_r}}{|\overrightarrow{a_1}\times\overrightarrow{a_r}|}\right|}. \tag{69}$$

Equation (69) is simplified further to equation (70) because $\left|\overrightarrow{a_2}\times\frac{\overrightarrow{a_1}\times\overrightarrow{a_r}}{|\overrightarrow{a_1}\times\overrightarrow{a_r}|}\right| = 1$.

$$\overrightarrow{dF_{12}} = \frac{\mu}{4\,\pi}\frac{I_1\,I_2}{|\overrightarrow{r}|^2}\,dl\,dl\,(\overrightarrow{a_2}\times\overrightarrow{a_1}\times\overrightarrow{a_r}). \tag{70}$$

Equation (70) is rewritten as equation (71).

$$\overrightarrow{dF_{12}} = \frac{\mu}{4\,\pi}\frac{I_1\,I_2}{|\overrightarrow{r}|^2}\,(dl\,\overrightarrow{a_2}\times(dl\,\overrightarrow{a_1}\times\overrightarrow{a_r})). \tag{71}$$

Equation (71) is rewritten as equation (72).

$$\overrightarrow{dF_{12}} = \frac{\mu}{4\,\pi}\frac{I_1\,I_2}{|\overrightarrow{r}|^2}\,(\overrightarrow{dl_2}\times(\overrightarrow{dl_1}\times\overrightarrow{a_r})). \tag{72}$$

where $\overrightarrow{dl_1} = dl\,\overrightarrow{a_1}$ is the directional infinitesimal length of the current element $\overrightarrow{dI_1}$, and $\overrightarrow{dl_2} = dl\,\overrightarrow{a_2}$ is the directional infinitesimal length of the current element $\overrightarrow{dI_2}$. Equation (72) is rewritten in terms of the current elements as equation (73).

$$\overrightarrow{dF_{12}} = \frac{\mu}{4\,\pi}\frac{1}{|\overrightarrow{r}|^2}\,(\overrightarrow{dI_2}\times(\overrightarrow{dI_1}\times\overrightarrow{a_r})). \tag{73}$$

Equation (73) is an exact equivalent, in both magnitude and direction, to the well-known magnetic force law between two perpendicular filamentary current elements that are fully contained by one plane. Furthermore, it is possible to apply the discontinuity charge analysis that is performed in this section to derive the electric force between two parallel filamentary current elements that are fully contained by one plane, as in section (IV-C1). The analysis should be performed in the relative frame for a static observer that is watching the two current elements. For example, let $\overrightarrow{dI_1}$ and $\overrightarrow{dI_2}$ be two parallel filamentary current elements with currents that propagate in the same direction, then, figures (24 and 25) show the discontinuity charges and the electric fields surrounding $\overrightarrow{dI_2}$ due to the existence of $\overrightarrow{dI_1}$ at two different positions: (1) where $|\overrightarrow{a_1}\times\overrightarrow{a_r}| = 1$ and (2) where $|\overrightarrow{a_1}\times\overrightarrow{a_r}| = 0$, respectively. By performing an analysis similar to the one provided in this section, equation (46) is obtained. The details for deriving equation (46) using the discontinuity charge analysis are omitted from this paper to avoid redundancy.

*3) Two Perpendicular Filamentary Current Elements on Two Perpendicular Planes:* This section derives the electric force law between two perpendicular filamentary current elements that are fully contained by two perpendicular planes. One element completely lies on one plane, while the other element completely lies on the second plane. The force law is derived for case (5) and case (6) of the six cases mentioned earlier, and then the general force law for two perpendicular elements contained by two perpendicular planes is found.

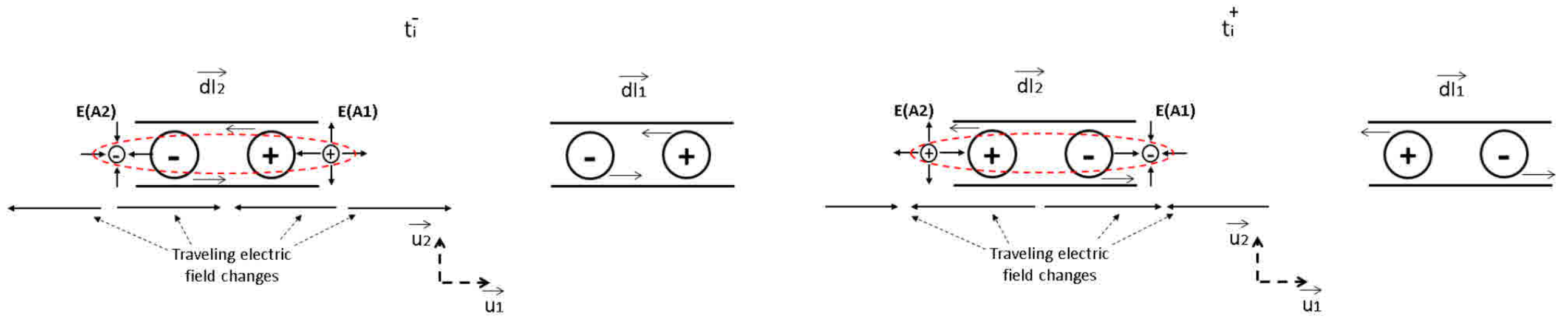

(a) (b)

Figure 25: Shows the discontinuity charges surrounding the current element $\overrightarrow{dI_2}$ due to the existence of current element $\overrightarrow{dI_1}$ at time $t_i$ for case (2), where $\overrightarrow{dI_1}$ and $\overrightarrow{dI_2}$ are parallel to each other and have currents that propagate in the same direction, on the negative $\overrightarrow{u_1}$-axis. (a) The discontinuity charges with their electric fields surrounding $\overrightarrow{dI_2}$ at time $t_i^-$. (b) The discontinuity charges with their electric fields surrounding $\overrightarrow{dI_2}$ at time $t_i^+$. $E(A_i)$ is the electric field at location $A_i$.

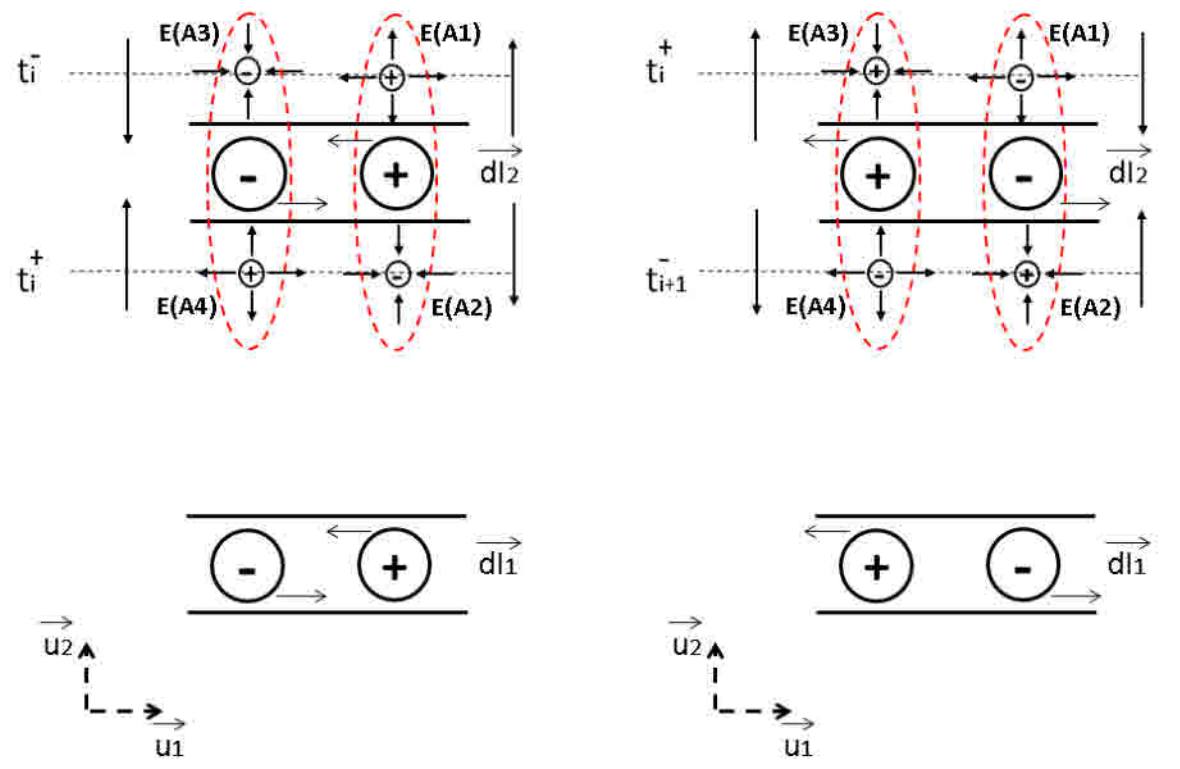

(a) (b)

Figure 24: Shows the discontinuity charges surrounding the current element $\overrightarrow{dI_2}$ due to the existence of current element $\overrightarrow{dI_1}$ at time $t_i$ for case (1), where $\overrightarrow{dI_1}$ and $\overrightarrow{dI_2}$ are parallel to each other and have currents that propagate in the same direction, on the negative $\overrightarrow{u_1}$-axis. (a) The discontinuity charges with their electric fields surrounding $\overrightarrow{dI_2}$ at time $t_i^-$. (b) The discontinuity charges with their electric fields surrounding $\overrightarrow{dI_2}$ at time $t_i^+$. $E(A_i)$ is the electric field at location $A_i$.

For case (5), let $\overrightarrow{dI_1}$ and $\overrightarrow{dI_2}$ be two perpendicular current elements, fully contained in two perpendicular planes, as shown in figure (26). The electric force on $\overrightarrow{dI_2}$ due to the existence of $\overrightarrow{dI_1}$ is computed by analyzing the electric force generated on the moving charges of $\overrightarrow{dI_2}$ due to the net relative current charge and due to the discontinuity charges surrounding $\overrightarrow{dI_2}$.

For the net relative charge of currents, the charges in $\overrightarrow{dI_2}$ see the positive and negative charges of $\overrightarrow{dI_1}$ cross the space of $\overrightarrow{dI_1}$ at the same speed all of the time. Thus, the same number of positive and negative charges crossed the space of $\overrightarrow{dI_1}$. Then, the net relative charge is zero. Therefore, the net generated force due to the relative current charge is zero in these two cases.

For the surrounding discontinuity charges, the current element $\overrightarrow{dI_2}$ has a positive discontinuity charge on one side and a negative discontinuity charge on the other side at all times, as shown in figure (27). This situation is similar to case (4) in section (IV-C2). The positive discontinuity charge produces a repulsive force on the positive charge under effect of $\overrightarrow{dI_2}$ and produces an attractive force of the same amount on the negative charge under effect of $\overrightarrow{dI_2}$. Meanwhile, the negative discontinuity charge produces a repulsive force on the negative charge under effect of $\overrightarrow{dI_2}$ and produces an attractive force of the same amount on the positive charge under effect of $\overrightarrow{dI_2}$. These forces are in opposite directions and are perpendicular to the direction of motion for the charges of $\overrightarrow{dI_2}$. These charges are not permitted to leave their filamentary current element. Given that, these charges push the filamentary current element by these forces. However, these forces have the same amount but in opposite directions; therefore, the net force applied on $\overrightarrow{dI_2}$ at any moment is zero, refer to equation (74).

$$\overrightarrow{dF_{12}} = 0. \tag{74}$$

For case (6), the two perpendicular filamentary current elements are now contained by one plane similar to case (4) when $\overrightarrow{a_1} \times \overrightarrow{a_r} = 0$. In this case, the net force applied on $\overrightarrow{dI_2}$ at any moment is zero as described in equation (74).

Using the result for case (5) and case (6), a general expression is written for the electric force law for two perpendicular current elements that are contained by two perpendicular planes at arbitrary positions. This expression uses equation (73) from section (IV-C2) because $\overrightarrow{a_2} \times \overrightarrow{a_1} \times \overrightarrow{a_r} = 0$ all the time for these cases. Thus, equation (73) evaluates to zero when applied for these two cases. Equation (73) is used to express equation (74). Equation (73) is an exact equivalent, in both magnitude and direction, to the well-known magnetic force law between two perpendicular current elements contained by two perpendicular planes.

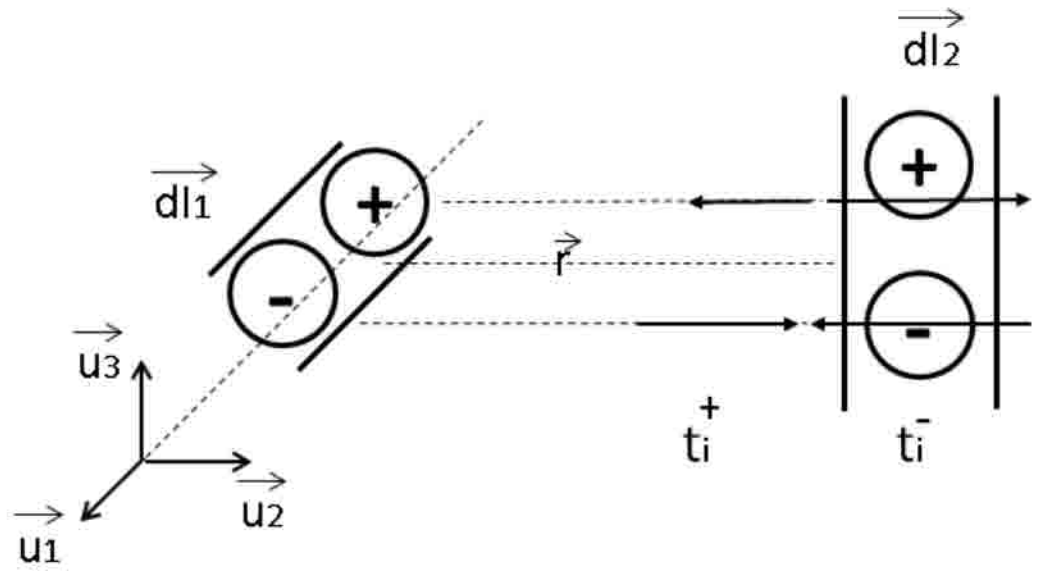

Figure 26: Shows case (5) for two perpendicular current elements $\overrightarrow{dI_1}$ and $\overrightarrow{dI_2}$ that are contained by two perpendicular planes. The generated pattern of changes in the electric field spreading in the space due to the movement of charges in a current element is shown at $\overrightarrow{dI_2}$ for moments $t_i^-$ and $t_i^+$.

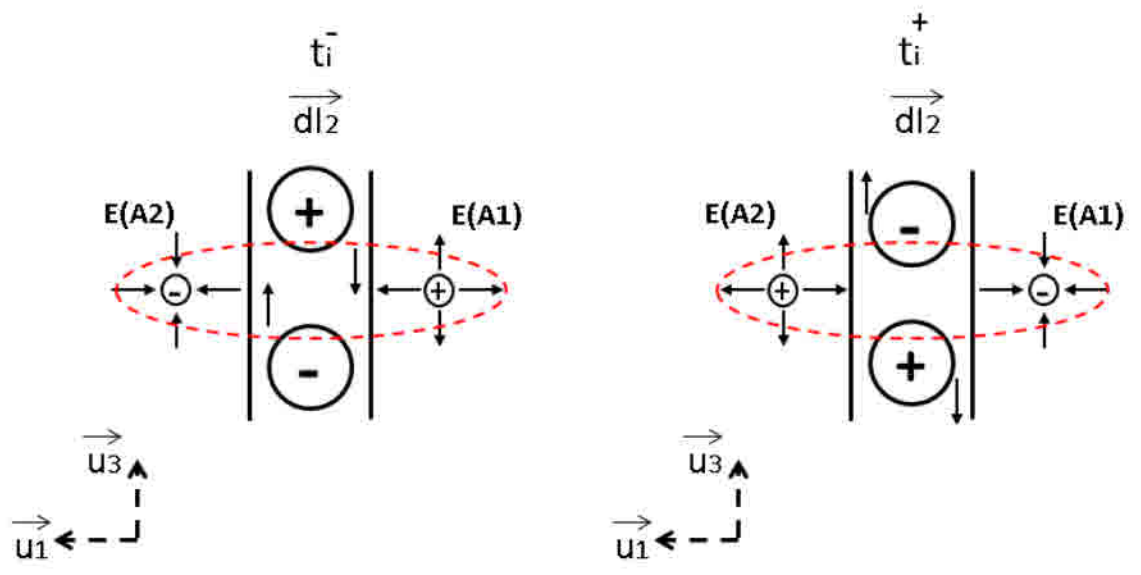

(a) (b)

Figure 27: Shows the discontinuity charges surrounding the current element $\overrightarrow{dI_2}$ due to the existence of current element $\overrightarrow{dI_1}$ at time $t_i$. Two perpendicular current elements $\overrightarrow{dI_1}$ and $\overrightarrow{dI_2}$ that are contained by two perpendicular planes. The currents propagate in the positive $\overrightarrow{u_1}$-axis and the negative $\overrightarrow{u_3}$-axis in $\overrightarrow{dI_1}$ and $\overrightarrow{dI_2}$, respectively. (a) The discontinuity charges with their electric fields surrounding $\overrightarrow{dI_2}$ at time $t_i^-$. (b) The discontinuity charges with their electric fields surrounding $\overrightarrow{dI_2}$ at time $t_i^+$. $E(A_i)$ is the electric field at location $A_i$.

*4) Deriving the Biot-Savart Law:* In this section, the general magnetic force law and Biot-Savart law between two filamentary current elements are derived using the electric force concept. This derivation is provided as a proof for the following second theory.

**Theory (2)**: The magnetic force between two filamentary current elements is an electric force in its origin. This force is the result of the electrical interaction between the current charges moving at the speed of light due to the relative charges of the current elements and due to the discontinuity charges produced by the pattern of the discontinuous electric field generated by the movement of the electric charges in the current elements.

**Proof**: Let $\overrightarrow{dI_1}$ and $\overrightarrow{dI_2}$ be two infinitesimal filamentary current elements in the space, as defined in equations (75-76).

$$\overrightarrow{dI_1} = I_1 dl\, \overrightarrow{a_1}, \tag{75}$$

$$\overrightarrow{dI_2} = I_2 dl\, \overrightarrow{a_2}. \tag{76}$$

where $I_1$ and $I_2$ are the amounts of current flowing through $\overrightarrow{dI_1}$ and $\overrightarrow{dI_2}$, respectively. $dl$ is the infinitesimal length of the current elements. $\overrightarrow{a_1}$ and $\overrightarrow{a_2}$ are the propagation directions for the current flows in $\overrightarrow{dI_1}$ and $\overrightarrow{dI_2}$, respectively. Let $\overrightarrow{r}$ be the distance vector between $\overrightarrow{dI_1}$ and $\overrightarrow{dI_2}$, as represented in equation (77).

$$\overrightarrow{r} = \overrightarrow{p_2} - \overrightarrow{p_1} = |\overrightarrow{r}|\, \overrightarrow{a_r}. \tag{77}$$

where $\overrightarrow{p_1}$ and $\overrightarrow{p_2}$ are the positions for $\overrightarrow{dI_1}$ and $\overrightarrow{dI_2}$, respectively. $|\overrightarrow{r}|$ and $\overrightarrow{a_r}$ are the magnitude and the unit vector for $\overrightarrow{r}$. Let $\overrightarrow{a_n}$ be the normal unit vector for the plane that contains $\overrightarrow{dI_1}$ and the vector $\overrightarrow{r}$, as computed in equation (78).

$$\overrightarrow{a_n} = \frac{\overrightarrow{a_1} \times \overrightarrow{a_r}}{|\overrightarrow{a_1} \times \overrightarrow{a_r}|}. \tag{78}$$

Let $\overrightarrow{u_1}$, $\overrightarrow{u_2}$, and $\overrightarrow{u_3}$ be orthonormal vectors that define a 3D space for $\overrightarrow{dI_1}$ and $\overrightarrow{dI_2}$, as computed in equations (79-81).

$$\overrightarrow{u_1} = \overrightarrow{a_1}, \tag{79}$$

$$\overrightarrow{u_2} = \frac{\overrightarrow{a_n} \times \overrightarrow{a_1}}{|\overrightarrow{a_n} \times \overrightarrow{a_1}|}, \tag{80}$$

$$\overrightarrow{u_3} = \overrightarrow{a_n}. \tag{81}$$

Then, the direction vectors for current elements $\overrightarrow{a_1}$ and $\overrightarrow{a_2}$ are rewritten in terms of the orthonormal vectors of the 3D space as in equations (82-83).

$$\overrightarrow{a_1} = \overrightarrow{u_1}, \tag{82}$$

$$\overrightarrow{a_2} = \cos(\varphi)\sin(\theta)\,\overrightarrow{u_1} + \sin(\varphi)\sin(\theta)\,\overrightarrow{u_2} + \cos(\theta)\,\overrightarrow{u_3}. \tag{83}$$

where $\varphi$ and $\theta$ are the polar and azimuth angles, respectively. The current element $\overrightarrow{dI_2}$ is rewritten as in equation (84).

$$\overrightarrow{dI_2} = I_2\, dl\, (\cos(\varphi)\sin(\theta)\,\overrightarrow{u_1} + \sin(\varphi)\sin(\theta)\,\overrightarrow{u_2} + \cos(\theta)\,\overrightarrow{u_3}). \tag{84}$$

The current elements with their orthonormal components are shown in figure (7).

The electric force that affects the current element $\overrightarrow{dI_2}$ due to the existence of $\overrightarrow{dI_1}$ is found by computing the electric force for each orthonormal component of $\overrightarrow{dI_2}$.

For the $\overrightarrow{dI_2}$ component along $\overrightarrow{u_1}$, the current element $\overrightarrow{dI_1}$ and this component are parallel to each other. Then, the electric force is generated due to the relative charges of the currents. This force is computed according to equation (46) and defined as in equation (85).

$$\overrightarrow{dF_{12\,u1}} = \frac{\mu}{4\pi}\frac{I_1 I_2}{|\overrightarrow{r}|^2}\, dl\, dl\, \cos(\varphi)\sin(\theta)\,(\overrightarrow{u_1} \times (\overrightarrow{a_1} \times \overrightarrow{a_r})). \tag{85}$$

where $\overrightarrow{dF_{12\,u1}}$ is the electric force applied on the $\overrightarrow{dI_2}$ component along $\overrightarrow{u_1}$ due to the existence of $\overrightarrow{dI_1}$.

For the $\overrightarrow{dI_2}$ component along $\overrightarrow{u_2}$, the current element $\overrightarrow{dI_1}$ and this component are perpendicular to each other, and they are fully contained by one plane. Then, the electric force is generated due to the discontinuity charges surrounding this component of $\overrightarrow{dI_2}$. These discontinuity charges are produced by the pattern of the discontinuous electric field generated due to the movement of the electric charges of the current element $\overrightarrow{dI_1}$. This force is computed according to equation (73) and defined as in equation (86).

$$\overrightarrow{dF_{12\,u2}} = \frac{\mu}{4\pi}\frac{I_1\,I_2}{|\overrightarrow{r}|^2}\,dl\,dl\,\sin(\varphi)\sin(\theta)\,(\overrightarrow{u_2}\times(\overrightarrow{a_1}\times\overrightarrow{a_r})). \quad (86)$$

where $\overrightarrow{dF_{12\,u2}}$ is the electric force applied on the $\overrightarrow{dI_2}$ component along $\overrightarrow{u_2}$ due to the existence of $\overrightarrow{dI_1}$.

For the $\overrightarrow{dI_2}$ component along $\overrightarrow{u_3}$, the current element $\overrightarrow{dI_1}$ and this component are perpendicular to each other, and they are contained by two perpendicular planes. Then, there is no electric force generated due to the relative charges of the current element or the discontinuity charges surrounding this component of $\overrightarrow{dI_2}$. This force is evaluated to zero according to equation (74). However, $\overrightarrow{u_3}\times\overrightarrow{a_1}\times\overrightarrow{a_r} = 0$, and then this force is defined as in equation (87).

$$\overrightarrow{dF_{12\,u3}} = \frac{\mu}{4\pi}\frac{I_1\,I_2}{|\overrightarrow{r}|^2}\,dl\,dl\,\cos(\theta)\,(\overrightarrow{u_3}\times(\overrightarrow{a_1}\times\overrightarrow{a_r})). \quad (87)$$

where $\overrightarrow{dF_{12\,u3}}$ is the electric force applied on the $\overrightarrow{dI_2}$ component along $\overrightarrow{u_3}$ due to the existence of $\overrightarrow{dI_1}$, and it is zero at all times.

The total electric force that affects $\overrightarrow{dI_2}$ due to the existence of $\overrightarrow{dI_1}$, denoted as $\overrightarrow{dF_{12}}$, is defined by equation (88).

$$\overrightarrow{dF_{12}} = \overrightarrow{dF_{12\,u1}} + \overrightarrow{dF_{12\,u2}} + \overrightarrow{dF_{12\,u3}}. \quad (88)$$

By substituting the values of $\overrightarrow{dF_{12\,u1}}$, $\overrightarrow{dF_{12\,u2}}$, and $\overrightarrow{dF_{12\,u3}}$, equation (88) is rewritten as equation (89).

$$\overrightarrow{dF_{12}} = \frac{\mu}{4\pi}\frac{I_1\,I_2}{|\overrightarrow{r}|^2}\,dl\,dl\,((\cos(\varphi)\sin(\theta)\,\overrightarrow{u_1} + \sin(\varphi)\sin(\theta)\,\overrightarrow{u_2} + \cos(\theta)\,\overrightarrow{u_3})\times(\overrightarrow{a_1}\times\overrightarrow{a_r})). \quad (89)$$

By using $\overrightarrow{a_2}$, equation (89) is rewritten as equation (90).

$$\overrightarrow{dF_{12}} = \frac{\mu}{4\pi}\frac{I_1\,I_2}{|\overrightarrow{r}|^2}\,dl\,dl\,(\overrightarrow{a_2}\times(\overrightarrow{a_1}\times\overrightarrow{a_r})). \quad (90)$$

By arranging the terms, equation (90) is rewritten as equation (91).

$$\overrightarrow{dF_{12}} = \frac{\mu}{4\pi}\frac{I_1\,I_2}{|\overrightarrow{r}|^2}\,(dl\,\overrightarrow{a_2}\times(dl\,\overrightarrow{a_1}\times\overrightarrow{a_r})). \quad (91)$$

Equation (91) is rewritten as equation (92).

$$\overrightarrow{dF_{12}} = \frac{\mu}{4\pi}\frac{I_1\,I_2}{|\overrightarrow{r}|^2}\,(\overrightarrow{dl_2}\times(\overrightarrow{dl_1}\times\overrightarrow{a_r})). \quad (92)$$

where $\overrightarrow{dl_1} = dl\,\overrightarrow{a_1}$ and $\overrightarrow{dl_2} = dl\,\overrightarrow{a_2}$. Equation (92) is an exact equivalent, in both magnitude and direction, to the well-known magnetic force law between two perpendicular filamentary current elements. Then, the virtual field that generates a force due to the existence of $\overrightarrow{dI_1}$ at any point in the space is found by taking out the terms of the current element $\overrightarrow{dI_2}$ from equation (92). This virtual field is computed via equation (93).

$$\overrightarrow{dB_1(r)} = \frac{\mu}{4\pi}\frac{I_1}{|\overrightarrow{r}|^2}\,(\overrightarrow{dl_1}\times\overrightarrow{a_r}). \quad (93)$$

where $\overrightarrow{dB_1(r)}$ is the virtual field generated by the infinitesimal current element $\overrightarrow{dI_1}$. Equation (93) is the Biot-Savart law, and this virtual field is known as the magnetic field.

**Done.**

The proof starts by defining a 3D space that contains the two current elements $\overrightarrow{dI_1}$ and $\overrightarrow{dI_2}$. The 3D space is formed by two perpendicular planes: the plane that contains $\overrightarrow{dI_1}$ and $\overrightarrow{r}$, and the other plane is its perpendicular. The current element $\overrightarrow{dI_2}$ is analyzed to find its three orthonormal components. Then, for each component, the electric force is computed following the analysis provided in sections (IV-C1,IV-C2, and IV-C3). The total force on $\overrightarrow{dI_2}$ due to the existence of $\overrightarrow{dI_1}$ is computed by combining the forces on the three orthonormal components. This total force is an exact equivalent, in both magnitude and direction, to the well-known magnetic force law between two filamentary current elements. The virtual field that generates this force is computed by taking out $\overrightarrow{dI_2}$ from the total force law described in equation (92). The derived equation for the virtual field is an exact equivalent to the Biot-Savart law for the magnetic field. This theory helps in providing an explanation for the properties of the known magnetic force law between two filamentary current elements, e.g., the relationship between the force and current propagation directions, as well as the relationship between the force and the displacement vector between the current elements. Moreover, according to this theory, electric charges of current elements interact with each other through either relative charges or discontinuity charges. The electric force between relative charges as well as between a current charge and a discontinuity charge satisfies Newton's third law. The forces exerted on current charges allow the charges to produce either a non-zero or zero net force on the containing infinitesimal current element. The produced net force is non-zero on the current element when the positive and negative charges push the current element in the same direction. However, the net force is zero when these charges push the current element in opposite directions thereby canceling each other or when the exerted forces on the current charges are completely along the direction of movement for the charges. The provided analysis is developed for an infinitesimal filament element. The total applied force on a larger filament or larger volume is found by summing the contributions from all the infinitesimal filament elements comprising the larger one.

## V. Conclusion

This paper has presented two new theories and a new current representation to explain the magnetic force between two filamentary current elements using electric forces. The first theory states that a current has an electric charge relative to its moving observer. The current relative charge is zero, negative or positive depending on the motion of its observer. The second theory states that the magnetic force between two filamentary current elements is a result of electric force interactions between current charges. The new current representation characterizes an electrically neutral current with equal amounts of positive and negative point charges moving in opposite directions at the speed of light. This representation is referred to as light-speed current representation. As a proof, the exact formulas for the magnetic force law between two current elements and the Biot–Savart law are derived using the electric force law between electric charges. The derivation process depends on analyzing the relative charge of current elements and the discontinuity charges generated due to position-switching between positive and negative charges of current elements. The existence of discontinuity charge effect in vacuum raises questions about the nature of the electric force and the nature of the electric charge. The investigation to answer these questions may need to be conducted in connection with other explorations to the nature of gravity and nuclear forces within a unified framework, e.g., string theory. Such an investigation is out of the scope of this work and is better suited for future research. These theories help in explaining the properties of the magnetic force and why this force occurs with moving charges and not with static ones. These new theories and the light-speed current representation have been developed to help in unifying the concepts of magnetism and electricity. This unification may have important engineering benefits that may allow engineers to enhance the electrical properties for materials and to design new algorithms for computational electromagnetism.